\documentclass[aps,prd,groupedaddress,preprint,eqsecnum,nofootinbib]{revtex4}
\usepackage{graphicx,epsf,amssymb,amsbsy,amsfonts,amssymb,amsmath,appendix}
\usepackage[papersize={8.5in,11in},hmargin=3cm,vmargin = 2cm]{geometry}
\usepackage{xcolor}
\usepackage[hyperfootnotes=false]{hyperref}

\newcommand{\IR}{\relax{\rm I\kern-.18em R}}
\def\blfootnote{\xdef\@thefnmark{}\@footnotetext}

\newcommand{\IC}{{\relax\hbox{$\inbar\kern-.3em{\rm C}$}}}
\newcommand{\IP}{\relax{\rm I\kern-.18em P}}
\newcommand{\IZ}{\relax\ifmmode\mathchoice
{\hbox{\cmss Z\kern-.4em Z}}{\hbox{\cmss Z\kern-.4em Z}}
{\lower.9pt\hbox{\cmsss Z\kern-.4em Z}} {\lower1.2pt\hbox{\cmsss
Z\kern-.4em Z}}\else{\cmss Z\kern-.4em Z}\fi} \font\cmss=cmss10
\font\cmsss=cmss10 at 7pt
\newcommand{\inbar}{\,\vrule height1.5ex width.4pt depth0pt}
\def\th{\theta}
\def\expp#1{{\rm exp}\big ( {#1} \big )}
\def\uu{^}
\def\llo{_}
\def\sqd{^2}
\def\hh{{\frac{1}{2}}}
\def\fakeindent{\hskip .2in}
\def\llsk{\fakeindent}

\definecolor{Sorbus}{rgb}{0.996094, 0.429688, 0.0273438}

\definecolor{FrenchRose}{rgb}{0.96875, 0.292969, 0.5625}

\definecolor{Violet}{rgb}{0.5,0,1}

\definecolor{red}{rgb}{1,0,0}

\usepackage{ulem}
\def\redlowdash{{\color{red}{\rule[-0.5ex]{2pt}{0.4pt}}}}
\def\reduline{\bgroup \markoverwith \redlowdash \ULon}
\def\shc{\bgroup \markoverwith \redlowdash \ULon}

\definecolor{Blue}{rgb}{0, 0.429688, 0.8}

\begin{document}
\hfuzz 12pt

\preprint{IPMU-12-0105, SITP 12/19}
\title{Light States in Chern-Simons Theory Coupled to Fundamental Matter\vspace{0.5in}}

%
\author{Shamik Banerjee$^{a}$, Simeon Hellerman$^{b}$, Jonathan Maltz$^{a}$, Stephen H. Shenker$^{a}$ }

\affiliation{$^{a}$Stanford Institute for Theoretical Physics and Department of
Physics, Stanford University,
Stanford, CA 94305-4060, USA\\$^{b}$Kavli IPMU, The University of Tokyo Kashiwa, Chiba 277-8583, Japan\vspace{0.5in}} 

\email{bshamik@stanford.edu}\email{ simeon.hellerman.1@gmail.com} \email{jdmaltz@stanford.edu} \email{sshenker@stanford.edu}





\begin{abstract}
Motivated by developments in vectorlike holography, we study SU(N) Chern-Simons theory coupled to matter fields in the fundamental representation on various spatial manifolds.  
On the spatial torus $T\uu 2$, we find  light states at small `t Hooft coupling $\lambda=N/k$, where $k$ is the Chern-Simons level, taken to be large.
In the free scalar theory the gaps are of order $\sqrt{\lambda}/N$ and in the critical scalar theory and the free fermion theory they are of order $\lambda/N$.  The entropy of these states grows like $ N \log(k)$.   We briefly consider spatial surfaces of higher genus.  Based on results from pure Chern-Simons theory,
it appears that there are light states with entropy that grows even faster,  like $N^2 \log(k)$.  This is consistent with the log of the partition function on the three sphere $S\uu 3$, which also behaves like $N^2 \log(k)$.   These light states require bulk dynamics beyond standard Vasiliev higher spin gravity to explain them. 
\end{abstract}

\maketitle

\tableofcontents

\section{Introduction}

There has been increasing interest in recent years in ``vectorlike" examples \cite{Klebanov:2002ja, Sezgin:2002rt, Petkou:2003zz, Sezgin:2003pt,Girardello:2002pp, Giombi:2009wh, Giombi:2010vg,Das:2003vw,Koch:2010cy, Douglas:2010rc,Giombi:2011ya,Shenker:2011zf,Giombi:2011kc,Aharony:2011jz,Maldacena:2012sf,Maldacena:2011jn} of holography that involve dynamical fields that transform in the fundamental (rather than the adjoint) representation of a symmetry group such as $SU(N)$ at large $N$. In 3+1 bulk dimensions the bulk dynamics is described by Vasiliev higher spin gravity\cite{Vasiliev:1992av,Vasiliev:1995dn,Vasiliev:1999ba,Vasiliev:2003ev,Sezgin:2002ru,Sezgin:2003pt} .  In these systems the bulk higher spin fields correspond to singlets under the symmetry group.\footnote{See \cite{Anninos:2011ui} for a very interesting proposal for a higher spin dS/CFT duality.}   A consistent implementation of AdS/CFT requires a boundary theory that is local, and so has a stress tensor.  In a local theory we can consistently truncate a global symmetry to the singlet sector only by a local procedure, such as
promoting the symmetry to a gauge symmetry and implementing the singlet constraint by local gauge interactions.  On the other hand we do not want dynamical gauge fields that have nontrivial local gauge invariant operators that are dual to extra  ``stringy" states in the bulk beyond those conjectured by the duality.   In $2+1$ boundary dimensions there is a natural candidate discussed in the literature \cite{Giombi:2011kc} for the gauge system that does not have nontrivial local gauge dynamics, the Chern-Simons theory.   We will study this proposal in this paper, focusing on the case with gauge group $SU(N)$ and matter fields that are in the fundamental $N$ dimensional vector representation.  There is an important parallel development in terms of  $1+1$ dimensional boundary systems, involving a $W_{N}$ boundary CFT that we will mention in the discussion\cite{Gaberdiel:2010pz,Gaberdiel:2011wb,Gaberdiel:2011nt,Ahn:2011pv,Chang:2011mz,Papadodimas:2011pf}.

But Chern-Simons theories have nontrivial dynamics and extra states on topologically nontrivial manifolds, and we shall see that the nontrivial Chern-Simons dynamics remain active when
coupled to matter as well. Since gauge/gravity dualities must make sense on any boundary manifold, these new states must be part of the full dual gravitational dynamics.

In this paper we take a first step towards understanding this situation by analyzing the Chern-Simons theory coupled to fundamental scalars and fermions on higher genus spatial surfaces, especially the torus $T^2$.   

We begin with the warm-up example of a massive scalar field. This is not a conformal field theory (CFT) and is not dual to the bulk higher spin theory. On the other hand the mass acts as a control parameter which makes the analysis less complicated when the mass squared is of the same order of or is large compared to the inverse size of the torus. In that case we can integrate out the scalar field and the low energy theory is pure Chern-Simons theory perturbed by the operators obtained by integrating out the scalar field. We can easily state the result of our analysis. In the weak-coupling or large $k$ limit the splitting of the exact zero energy states of the pure Chern-Simons theory on torus is of order $\frac{1}{k} = \frac{\lambda}{N}$, where we have defined the 't Hooft coupling $\lambda$ as, $\lambda=\frac{N}{k}$.  

The other case we study is that of a free massless scalar field\footnote{To be precise we study the interacting fixed point theory with $\phi^6$ coupling of order $\lambda^2$ discovered in \cite{Aharony:2011jz}.}. This theory is dual to a bulk Vasiliev theory on locally $AdS_{4}$ space whose asymptotic boundary has the structure of $T^{2}\times R^{1}$.\footnote{This can be constructed by periodic identification of space-like field theory coordinates in the Poincare patch of $AdS_{4}$.} This is more complicated than the massive scalar because of the presence of the scalar zero mode. This is an approximate zero mode but it can have arbitrarily small energy and so it does not decouple from the low energy dynamics.   We study the low energy spectrum by reducing this system to an effective quantum mechanics.  We analyze this quantum mechanics and find for the $U(1)$ system the gap is $\sim \frac{1}{\sqrt{k}}$.  For the $SU(N)$ system the gap  is of order $\frac{\sqrt{\lambda}}{N}$ where the `t Hooft coupling $\lambda = N/k$. and vanishes in the large-$N$ limit.  So the bulk higher spin theory must have extremely low energy states in the classical, and small $\lambda$ limits.
These light states do not correspond to any apparent excitations of the Vasiliev fields.
They are closely analogous to the light states found in the $W\llo N$ theory
\cite{Gaberdiel:2010pz, Gaberdiel:2011zw, Gaberdiel:2011aa}.

 The critical, interacting,  $SU(N)$ scalar  theory, dual to Vasiliev gravity with a different bulk scalar boundary condition,  has parametric behavior similar to the free massive  scalar because it is gapped on the torus.  The free fermion system behaves in the same fashion.
The entropy corresponding to these light states is $S \sim N \log(k)$.

We also consider, briefly, the case where the spatial slices are Riemann surfaces of
higher genus and,  based on  results in pure Chern-Simons theory,
we find an even larger entropy, $ S \sim N^2 \log(k)$ parametrically in $N$, than
in the case of genus 1.  While we focus on the torus case for concreteness, 
it seems that the higher-genus case portends an even more radical breakdown of
the bulk description in terms of pure Vasiliev gravity.

\section{Perturbative Chern-Simons Matter Theory On Torus}
Pure Chern-Simons theory on a general three manifold is an exactly solvable field theory for any $k$ \cite{Witten:1988hf}.  On a torus the space of states is determined by the conformal blocks of WZW conformal field theory \cite{Witten:1988hf}. However it is often useful to understand the theory in a semiclassical weak coupling expansion at large $k$.   The classical stationary points are flat connections.  On certain manifolds the flat connections are isolated and the semiclassical expansion is in principle straightforward.  On other manifolds, including tori, the flat connections form a moduli space which must be integrated over \footnote{ References \cite{Axelrod:1991vq} and \cite{Axelrod:1993wr} study pure Chern-Simons perturbation theory. They develop the perturbation theory based on the assumptions that the flat connection is isolated and the subgroup of the gauge group which leaves the flat connection invariant is discrete. These assumptions are violated if the spatial slice is a torus or any higher genus Riemann surface. A generalization of their method should be applicable to these cases and in general to any three manifold where there is a moduli space of flat connections. } \cite{Axelrod:1991vq,Axelrod:1993wr}.   An important approach to this problem is the canonical quantization method described in the classic paper \cite{Elitzur:1989nr} In In this approach the problem reduces to the quantization of the moduli space of flat connections on a spatial torus, which form a finite volume phase space.   The Hilbert space is finite dimensional and every state has exactly zero energy because the Hamiltonian of the Chern-Simons theory vanishes.   We shall follow the canonical approach to study the more complicated problem of Chern-Simons theory coupled to a scalar matter field $\phi_a$  in the fundamental representation of $SU(N)$.

 Chern-Simons gauge theory coupled to a matter field is not  a topological field theory and it has a non-vanishing Hamiltonian. The presence of the scalar field lifts the degeneracy of the flat connections.  This was first studied in  the pioneering work of Niu and Wen \cite{Wen:1990zza} which has important parallels to our work.   To see how the flat directions behave, let us write down the action of the theory. The action is given by


\begin{equation}
S = \frac{k}{4\pi} \int Tr(A\wedge dA + \frac{2}{3} A^{3}) + \int d^{3}x  \sqrt {-g} \ [ g^{{\mu}{\nu}}  (D_{\mu}{\phi})^{\dagger}(D_{\nu}{\phi}) - V(\phi^{\dagger}\phi) ]
\end{equation}
where $g_{\mu\nu}$ is the metric on the space-time, $V(\phi^{\dagger}\phi)$ is the potential, $\phi^{\dagger}\phi = \phi_a \phi_a$, and $D_{\mu}\phi = \partial_{\mu} \phi + iA_{\mu}\phi$ is the gauge covariant derivative acting on the scalar field. We take the space-time to be $T^{2}\times R^{1}$, where $R^{1}$ is the time direction. Since the Chern-Simons term is topological it does not contribute to the stress tensor of the theory. So the energy density is given by 
\begin{equation}
T^{00} \sim |D_{0}\phi|^{2} + |D_{i}\phi|^{2} + V(\phi^{\dagger}\phi)
\end{equation}
Before we proceed farther it is useful to choose a gauge. For our purpose, $A_{0} = 0$ is the convenient gauge choice. In this gauge the energy density becomes,
\begin{equation}\label{xxx}
T^{00} \sim |\dot\phi|^{2} + |D_{i}\phi|^{2} + V(\phi^{\dagger}\phi)
\end{equation}
The allowed field configurations also have to satisfy the Gauss's law constraint,
\begin{equation}
\frac{ k}{8\pi}\frac {\epsilon^{ij}}{\sqrt h} \ F^{a}_{ij} \ = \ i(\phi^{\dagger}T^{a}\dot\phi - \dot\phi^{\dagger}T^{a}\phi)
\end{equation}
where $h_{ij}$ is the metric tensor on the spatial torus and $\epsilon^{ij}$ is the completely antisymmetric symbol with $\epsilon^{12}= - \epsilon^{21} = 1$. In the following discussion we shall specialize to the case where $V(\phi^{\dagger}\phi) = M^{2}\phi^{\dagger}\phi$.  

It follows from the expression of the energy density (\ref{xxx}) that classically the lowest energy field configurations are those for which, $\dot\phi = 0$, $D_{i}\phi = 0$ and $V(\phi^{\dagger}\phi) = 0$. The simultaneous solutions of these equations also have to satisfy the Gauss's law constraint. Since $\dot\phi = 0$, it follows from the Gauss's law constraint that the spatial components of the gauge field are components of a flat connection on the torus.  Now one can solve the remaining two equations subject to the constraint that the gauge field appearing in the covariant derivative is flat. If $M\neq 0$ then the only solution is $\phi = 0$ and so the classical lowest (zero) energy field configurations are flat connections on the torus. When $M=0$, one can show that the solution is $\phi = 0$, for almost every flat connection except for those whose holonomies lie in a $SU(N-1)$ subgroup of the gauge group $SU(N)$. Although the constant mode of the massless scalar field is not an exact zero mode when coupled to gauge fields, it can have arbitrarily small energy depending on the choice of the flat connection and so it does not decouple from the low energy dynamics. This will play an important role in the following discussion. 

In the quantum theory the flat connection  degeneracy is lifted by the scalar field. To show this, we can choose a particular flat connection and expand around that. The gauge field can be decomposed as,
\begin{equation}
A = A^{f} + \frac{1}{\sqrt k} a
\end{equation}
where $A^{f}$ is a flat connection and $a$ is gauge field fluctuation. Substituting this in the action we get a term of the form,
\begin{equation}
S\supset \int d^{3}x \sqrt {-g}  [(D_{\mu}^{f} \phi)^{\dagger} D^{f\mu}\phi + M^{2}\phi^{\dagger}\phi ] + O(\frac{1}{\sqrt k})
\end{equation}
where $D_{\mu}^{f} = \partial_{\mu} + iA_{\mu}^{f}$. In the weak coupling or large $k$ limit this is the leading piece of the action containing the scalar field. This is the action of a massive scalar field moving in the background of a flat connection.  For finite $M$ we can integrate out the scalars to get an effective potential for the flat connections\footnote{Terms with higher derivatives, like Yang Mills terms, are also induced.  Their effects  are suppressed at large $k$ .  For the abelian case this can be seen in the results of \cite{Wen:1990zza,  Gukov:2004id}}. The answer is given by\footnote{Please see the appendix for a detailed derivation of the formula and explanation of various terms.},
\begin{equation}\label{vaa}
V(A_{1}^{f},A_{2}^{f}) = - \frac{1}{\sqrt 2} \frac{area(T^{2})}{\pi^{\frac{3}{2}}}\sum_{(n,m)\neq(0,0)} \frac{M^{\frac{3}{2}}K_{\frac{3}{2}}(M|m\vec a + n\vec b|)}{|m\vec a + n\vec b|^{\frac{3}{2}}} Tr(W(a)^{m}W(b)^{n})
\end{equation}
which acts as an effective potential for the flat connections. In the above formula $W(a)$ and $W(b)$ are the holonomies of the flat connection along the $a$-cycle and $b$-cycle of the torus, respectively. It is easy to see that the effective potential has minima at $(A_{1},A_{2}) = (0,0)$ and its gauge copies. So quantum mechanically only the trivial flat connection is stable and one can quantize only around the trivial flat connection.\footnote{A very similar argument was used in \cite{Luscher:1982ma} for the case of pure Yang-Mills gauge theory on the three torus $T^{3}$.} Now what is the effect of the mass of the scalar field? It is expected that if the scalar field is heavy, i.e, $M^{2}\times area(T^{2}) \gg 1$, then the low energy theory should reduce to the pure Chern-Simons theory. In particular the effective potential on the moduli space should vanish as the mass tends to infinity. It is easy to see by studying the asymptotics of the modified Bessel function for large argument that this is indeed the case with the potential. 

So we have the following picture. The scalar field creates an effective potential on the moduli space of flat connections which push the connection towards the trivial one.\footnote{We would like to mention that this is not true in a supersymmetric theory. In a supersymmetric theory the effective potential obtained by integrating out the non-zero modes will vanish due to Bose-Fermi cancellation. So we can no longer say that the low energy wave functions are localized around the trivial flat connection and its gauge copies. The dimensionally reduced quantum mechanical model does not capture the complete low energy spectrum in the supersymmetric theory.} The effect of the scalar field or the value of the effective potential depends on its mass. In the regime $M^{2}\times area(T^{2}) \gg 1$, the potential decreases exponentially like $e^{-M\sqrt {area(T^{2})}}$ and in the limit of an infinitely massive scalar field we recover the pure Chern-Simons gauge theory. For a finite mass scalar, or in general, a matter theory with gap,  we can analyze the Chern Simons theory with (\ref{vaa}) as perturbation.  We will discuss this in the next section.

  In the case of a massless scalar the scalar zero mode cannot be integrated out,  but the low energy dynamics $(E \times \sqrt{area(T^{2})}\ll 1)$ can be determined by retaining only the constant modes and studying an effective quantum mechanics.  Because the gauge fields are confined to a small neighborhood of the zero gauge field we can ignore the compactness of the flat connection moduli space in formulating the quantum mechanics.   So to compute the low energy states of the theory we can diagonalize the Hamiltonian obtained from the dimensionally reduced theory.   The study of this quantum mechanics in both the $U(1)$ and $SU(N)$  cases will occupy most of what follows.  We will turn to it after discussing the gapped theories.

Finally we briefly discuss the behavior of the system when bosons are replaced with fermions 
in the the fundamental representation.  The analysis above for the scalars applies, except
that the sign of the potential in (\ref{vaa}) is reversed.  As a result, the
minimum of the potential occurs when the gauge field holonomies
on both cycles are diagonal and equal to $-1$.  Equivalently, and more simply,
we can treat the fermions as having Scherk-Schwarz boundary conditions
along both cycles, and the gauge field holonomies as being trivial.
In either description, the free fermions expanded around their true vacuum have an energy gap
of order ${\frac{1}{{\sqrt{{area(T^2)}}}}}$ and can be integrated out, even in the absence
of a mass; the analysis of the effective theory then follows exactly that of the 
massive scalar, with a mass of order ${\frac{1}{{\sqrt{{area(T^2)}}}}}$.

\section{Scalar Field With Mass}
In this section we shall study the case of a scalar field with mass, $ M$ in more detail.\footnote{In this section we shall take a square torus with sides of length $R$ we will often set to 1.} A massive scalar field with mass in the region $M\sim 1/R$ or $M\gg 1/R$ is in some sense simpler because we can integrate out the scalar field, if we are interested in states  with energies $\ll 1/R$. In that case we will be left only with the pure Chern-Simons gauge theory with corrections which come from integrating out the massive scalar field.   For simplicity let us examine the $U(1)$ case first.  At large $M$ (\ref{vaa}) becomes 
\begin{equation}
V(A_1, A_2) \sim - \exp(-MR)(\cos(A_1)+\cos(A_2))
\end{equation}
where $A_1$ and $A_2$ are the eigenvalues of the holonomies around the $1$ and $2$ cycles.   The Lagrangian of the  system becomes 
\begin{equation}
L = R^{2}\Big[\frac{ k}{8\pi}\,A_{1}\frac{dA_{2}}{dt} + \exp(-MR)(\cos(A_1)+\cos(A_2))\Big]
\end{equation}
We will argue that at low energies the gauge fields are close to zero, so we can expand the cosines.  The Lagrangian becomes
\begin{equation}
L = R^{2}\Big[\frac{ k}{8\pi}\,A_{1}\frac{dA_{2}}{dt} - \hh~\exp(-MR)(A_1^2 +A_2^2)\Big]
\end{equation}
$A_1$ and $A_2$ are canonically conjugate variables.  If we define
 $P = \frac{ k}{8\pi}A_{1}$ and $Q=A_{2}$, the canonical commutation relations are given by,
\begin{equation}
[Q,P] = i , \  [\phi,\pi] = [\phi^{*},\pi^{*}] = i
\end{equation}

If we measure all energies in units of $\frac{1}{R}$ then we can set $R=1$. The Hamiltonian is given by,
\begin{equation}
H =\hh ~ \exp(-M)[(\frac{8\pi}{k})^2P^{2} + Q^{2} ]
\label{qmmass}
\end{equation}
So the low lying states are described by a harmonic oscillator with $\hbar \sim 1/k$.  The low lying states have energies\footnote{Niu and Wen \cite{Wen:1990zza} emphasized that the $\exp(-M)$ dependence showed the stability of topological order. }
\begin{equation}
E_l = \exp(-M)(\frac{8 \pi}{ k})(l+\frac{1}{2})
\end{equation}
The spread of the holonomy angle in these states $\langle A_1^2 \rangle = \langle A_2^2 \rangle \sim 1/k$ so the gauge fields are localized around the origin at large $k$.  This effect is independent of $M$, for large enough $k$.  This localization justifies expanding the cosines.

The $SU(N)$ case is similar.  We can write down the effective potential on the moduli space of flat connections created by the massive scalar field at large $M$ as\footnote{Please see the appendix for the exact expression valid for all values of the mass parameter.}, 
\begin{equation}
V \sim \exp(-M)(\sum_i\cos(\alpha_i)+\cos(\beta_i))
\end{equation}
where $\alpha_i, \beta_i$ are the eigenvalues of $(A_{1}^{f},A_{2}^{f})$.  Each pair $\alpha_i, \beta_i$ are canonically conjugate and so expanding the cosines yields a set of $N-1$ harmonic oscillators with $\hbar \sim 1/k$.  There is a residual discrete part of the Gauss's Law constraint, the Weyl group.  Here this is just the permutation group $S_{N-1}$.   A careful analysis of the measure in the pure Chern-Simons system \cite{Elitzur:1989nr,Douglas:1994ex} shows that the eigenvalues should be treated as fermions\footnote{ We would like to thank Juan Maldacena  for pointing this out to us, correcting an error in the original version of this paper.  (This does not alter our substantive conclusions.)  The fermi surface plays an important role in \cite{Aharony:2012ns} }.   So we have $N-1$ fermionic particles in a harmonic oscillator potential.

 This system is easy to study.  The low lying gaps are 
\begin{equation}
\Delta \sim \frac {\expp{-M}} {k}
 = \frac {\expp{-M} \lambda} {N}
\end{equation}
Here we have used the `t Hooft coupling appropriate for Chern Simons systems, $\lambda = N/k$.     We note that the Fermi energy of this system is given by, $E_F = e^{-M}\lambda$. 
The entropy for temperature $\expp{-M} \gg T \gg E_F$ and $k$ large  can easily be computed semiclassically.  The answer is
\begin{equation}
\label{OurEntropyA}
S  \sim \log\left [ (T~ k~ \expp{+M})^{N-1}/(N-1)! \right ]
\sim (N-1)\log \left [ T~ \expp{+M} /\lambda \right ]
\end{equation}
Note that at fixed $N$ and large $k$ the entropy goes like $N \log(k)$, parametrically
the same as the pure Chern-Simons entropy.   Here we have performed the calculation
using the harmonic oscillator representation of the system.  This approximation
is valid when the range of holonomies explored is much less than $2\pi$, which  
corresponds to temperatures much less than $\exp{-M}$.  For temperatures
\begin{equation}
\expp{-M} \ll    T\ll 1\ ,
\end{equation}
all Chern-Simons states participate in the thermal ensemble.  

We can also consider a different limit where the temperature is much less than the Fermi energy of the system. In this temperature regime the entropy of the system is given by,
\begin{equation}
S \sim  e^{M}\frac{NT}{\lambda}
\end{equation} 
This is the leading term in the Sommerfeld expansion.

In accordance with expectations from effective field theory, we have seen that there is intermediate range of temperatures in which the matter and the Kaluza-Klein modes have
decoupled, and the system is well-approximated by pure Chern-Simons theory.  

The above analysis strictly applies when $M \gg 1$.  But when $M$ decreases to of order one the only thing that changes is the detailed shape of the potential.  It still is quadratic around $A=0$.  So we still have a harmonic oscillator description of the low lying states at large $k$. This applies to a general matter sector.  As long as the matter sector on the torus has a gap, the effective potential for the flat connections will be quadratic around the origin.     Interacting critical points, like the Wilson-Fisher fixed points (often called in this context the critical SU(N) model)  have a gap on the torus, so this analysis applies to these systems.

We expect this analysis to breakdown when the gap of the matter system is $\sim 1/k$.  The massless scalar is an example.  To treat this system we will have to retain the light degrees of freedom in our effective description.  We now turn to this task.

\section{Field Content Of The Low Energy Theory With A Massless Scalar}
The energy density (\ref{xxx}) depends on the derivatives of the scalar field. So a field configuration with a spatially varying scalar field will have at least an energy of order $\frac{1}{\sqrt {area(T^{2})}}$. So at an energy scale much below this we can integrate out these modes. The scalar field also has an approximate zero mode. The constant mode of the scalar field is an exact zero mode when the gauge field configuration is such that, $D_{i}\phi = 0$. It is easy to check that this has a non-zero solution for $\phi$ only when the gauge field is vanishing or flat with the holonomy lying in a $SU(N-1)$ subgroup of the $SU(N)$.  Away form this region of the moduli space of flat connections the constant mode of the scalar field is not an exact zero mode but it can have arbitrarily small energy and so we shall keep this mode in the low energy effective quantum mechanics


The story of the gauge field goes like this. The energy density does not depend on the derivatives of the gauge field. The standard kinetic and potential terms of the gauge field involving the squares of the electric and magnetic fields is absent in this case. So it appears that a field configuration with a very large magnetic field can have energy small compared to $\frac{1}{\sqrt {area(T^{2})}}$. But this changes once we take into account the Gauss's law constraint. If we switch on a magnetic field such that the left hand side of the Gauss's law constraint is a spatially varying quantity then the scalar field on the right hand side will also have to |depend on the spatial coordinates. This requires an energy of order $\frac{1}{\sqrt {area(T^{2})}}$. So we can safely neglect (integrate out) such modes of the gauge field. 

\subsection{Abelian Gauge Theory}
The first example that we shall study is an Abelian gauge field coupled to complex scalars. We shall define the theory on the space-time manifold $T^{2}\times R^{1}$. Since the modular property of the torus will not play any role in our analysis we take a square torus with side of length $R$. The metric on the torus is given by,
\begin{equation} 
dh^{2} = dx_{1}^{2} + dx_{2}^{2}
\end{equation}
where $x_{i}\sim x_{i} + R$. To obtain the low energy effective quantum mechanics we keep only the zero momentum modes of the gauge and matter fields in the Lagrangian and the reduced Lagrangian becomes,
\begin{equation}
L = R^{2} [ \ \frac{ k}{8\pi} \ A_{1}\frac{dA_{2}}{dt} + \frac{d\phi^{*}}{dt}\frac{d\phi}{dt} - \phi^{*}\phi \ (A_{1}^{2} + A_{2}^{2}) \ ]
\end{equation}
As we have argued in the previous sections this dimensionally reduced theory can capture the very low energy ($E\ll 1/R$) states in the theory. In the $A_{0} = 0$ gauge the residual gauge transformations are the time independent $U(1)$ rotations. The reduced Lagrangian has a global $U(1)$ rotation symmetry which acts on the scalar field. Since the physical states have to be gauge invariant we shall treat this global $U(1)$ rotation symmetry of the quantum mechanics as a gauge symmetry. So the physical states are those which are invariant under the rotation. 

If we measure all energies in units of $\frac{1}{R}$ then we can set $R=1$. The Hamiltonian is given by,
\begin{equation}
H = \pi^{*}\pi + \phi^{*}\phi [ (\frac{8\pi}{ k})^{2} P^{2} + Q^{2} ]
\end{equation}
where we have defined $P = \frac{ k}{8\pi}A_{1}$ and $Q=A_{2}$. $\pi$ and $\pi^{*}$ are momenta canonically conjugate to $\phi$ and $\phi^{*}$. The canonical commutation relations are given by ,
\begin{equation}
[Q,P] = i , \  [\phi,\pi] = [\phi^{*},\pi^{*}] = i
\end{equation}
 A scaling argument shows  that the total energy of the system is proportional to $\frac{1}{\sqrt k}$.
In fact this problem is easy to solve exactly. The total wave function can be written as a product $\Psi(\phi,\phi^{*},Q) = \chi(\phi,\phi^{*})\psi(Q)$, where $\chi$ is the rotationally invariant wave function in the matter sector and $\psi$ is the wave function of the gauge sector. The gauge sector is a one dimensional harmonic oscillator and the matter sector is an isotropic two dimensional harmonic oscillator where the Gauss's Law constraint imposes rotational invariance.  The energy eigenvalues are given by
\begin{equation}
E_{j,l} \sim (2j+1)\sqrt{(\frac{16 \pi}{k})(l+\frac{1}{2})}
\end{equation}
 So at weak coupling or large $k$ this energy is well below the Kaluza-Klein scale which is of $O(1)$, which justifies our neglect of spatially varying modes.
The wave functions in the quantum mechanics are all concentrated near the origin, $A_{1} = A_{2} =0$. The width of the wave function goes like, $<A_{1}^{2}> \ = \ <A_{2}^{2}>  \ \sim \frac{1}{k}$. This is very small in the large $k$ limit which justifies our neglect of the periodicity of the holonomies $A_{1}, A_{2}$. 

\subsection{SU(N) Gauge Theory}
Our next example is $SU(N)$ Chern-Simons gauge theory coupled to massless fundamental scalar matter \footnote{The quantum mechanics arising in SU(2) gauge theory can be solved exactly. Please see the appendix for the exact solution.  }.   In the quantum mechanics problem we can treat the variables $A_{1}^{f}$ and $A_{2}^{f}$ as effectively non-compact. 

As in the U(1) case we  need to treat the dynamics of the zero mode of the scalar field by incorporating it and the gauge field dynamics in an effective dimensionally reduced quantum mechanics.  Because of the localization of the gauge field we can ignore the compactness of the flat connection moduli space.  The Lagrangian is given by
\begin{equation}
L = \frac{k}{8{\pi}} Tr(A_{1}\frac{d}{dt} A_{2})  +  \frac{d\phi^{\dagger}}{dt}\frac{d\phi}{dt}  - {\phi}^{\dagger}(A_{1}^{2}+A_{2}^{2}){\phi}
\end{equation}
Here the variables $A_{1}$ and $A_{2}$ are two arbitrary $N\times N$ Hermitian matrices not necessarily commuting, and 
$\phi$ is a complex $N$-dimensional column vector transforming in the fundamental representation of $SU(N)$. The Lagrangian has a global symmetry under which,
\begin{equation}
A_{i} \rightarrow UA_{i}U^{\dagger}, \  \phi\rightarrow U\phi
\end{equation}
where $U$ is a $SU(N)$ matrix. The $SU(N)$ global symmetry of the Lagrangian is the remnant of the $SU(N)$ gauge symmetry of the original field theory and so we should treat this $SU(N)$ symmetry as a gauge symmetry. A state in the quantum mechanics will be called physical if it is invariant under  $SU(N)$ transformations. 

To write the Lagrangian in a more manageable form we express the gauge potentials in terms of generators of the $SU(N)$ group. 
\begin{equation}
A_{i} = A_{i}^{a} T^{a} , \\  a = 1,.......,N^{2}-1
\end{equation}
where $T^{a}$ are generators in the fundamental representation which satisfy the relation, $Tr(T^{a}T^{b}) = \frac{1}{2}\delta_{ab}$. In terms of this the Lagrangian can be written as,
\begin{equation}
L = \sum_{a=1}^{N^{2}-1}P^{a}\frac{dQ^{a}}{dt} + \sum_{i=1}^{N}\frac{d\phi_{i}^{*}}{dt}\frac{d\phi_{i}}{dt} - \frac{1}{2} \sum_{a,b=1}^{N^{2}-1}M^{ab}(\phi) (P^{a}P^{b} + \hbar^{2}Q^{a}Q^{b})
\end{equation}
where, $P^{a} = A_{1}^{a}$, $Q^{a} = \frac{ k}{16\pi} A_{2}^{a}$, $\hbar = \frac{16\pi}{k}$ and $M^{ab}(\phi) = \phi^{\dagger}\lbrace T^{a},T^{b}\rbrace \phi$. It is clear from the form of the Lagrangian that $(P^{a},Q^{a})$ are canonically conjugate variables\footnote{We have made a small change definitions here relative to the U(1) case}. The Hamiltonian can be written as,
\begin{equation}
H = \sum_{i=1}^{N} \pi_{i}^{*}\pi_{i} + \frac{1}{2} \sum_{a,b=1}^{N^{2}-1}M^{ab}(\phi) (P^{a}P^{b} + \hbar^{2}Q^{a}Q^{b})
\end{equation}  
where $(\phi^{*},\pi)$ and $(\phi,\pi^{*})$ are canonically conjugate. the canonical commutation relations are given by,
\begin{equation}
[Q^{a},P^{b}] = i \delta^{ab},   [\phi_{i},\pi_{j}^{*}] = [\phi_{i}^{*},\pi_{j}] = i \delta_{ij}
\end{equation}
and rest of the commutators are zero. We can define the creation and annihilation operators in the gauge sector as,
\begin{equation}
\beta^{a} = \frac{1}{\sqrt {2\hbar}} (P^{a} - i\hbar Q^{a}), \  \beta^{a\dagger} = \frac{1}{\sqrt {2\hbar}}(P^{a} + i\hbar Q^{a})
\end{equation}
which satisfy the commutation relation,
\begin{equation}
[\beta^{a},\beta^{b\dagger}] = \delta^{ab}
\end{equation}
and rest of the commutators are zero. In terms of these operators the Hamiltonian can be rewritten as,
\begin{equation}
H = \pi_{i}^{*}\pi_{i} + \omega^{2}\phi_{i}^{*}\phi_{i} + \hbar M^{ab}(\phi) \beta^{a\dagger}{\beta}^{b}
\end{equation}
where we have defined, $\omega^{2} = \frac{N\hbar}{2}\sim \frac{N}{k}= \lambda$. We can also define the following creation and annihilation operators in the scalar sector,
\begin{equation}
\alpha_{i} = \frac{1}{\sqrt{2\omega}} (\pi_{i} - i\omega \phi_{i}), \  \alpha_{i}^{\dagger} = \frac{1}{\sqrt{2\omega}} (\pi_{i}^{*} + i\omega \phi_{i}^{*})
\end{equation}
\begin{equation}
\bar\alpha_{i} = \frac{1}{\sqrt{2\omega}} (\pi_{i}^{*} - i\omega \phi_{i}^{*}), \  \bar\alpha_{i}^{\dagger} = \frac{1}{\sqrt{2\omega}} (\pi_{i} + i\omega\phi_{i})
\end{equation}
they satisfy the commutation relation,
\begin{equation}
[\alpha_{i},\alpha_{j}^{\dagger}] = \delta_{ij}, \  [\bar\alpha_{i}, \bar\alpha_{j}^{\dagger}] = \delta_{ij} 
\end{equation}
and rest of the commutators are zero. In terms of these creation and annihilation operators the Hamiltonian can be written as,
\begin{equation}
\begin{split}
H& = \omega(\alpha_{i}^{\dagger}\alpha_{i} + \bar\alpha_{i}^{\dagger}\bar\alpha_{i}) + \frac{\hbar}{2\omega} \beta^{a\dagger}\beta^{a} + N\omega \\
  & -\frac{\hbar}{2\omega}\biggl [\bar\alpha_{i}\lbrace T^{a},T^{b}\rbrace_{ij}\alpha_{j} + \alpha_{i}^{\dagger}\lbrace T^{a},T^{b}\rbrace_{ij}\bar\alpha_{j}^{\dagger} \\  
  & - \bar\alpha_{j}^{\dagger}\lbrace T^{a},T^{b}\rbrace_{ij} \bar\alpha_{i} - \alpha_{i}^{\dagger}\lbrace T^{a},T^{b}\rbrace_{ij} \alpha_{j}\biggr] \beta^{a\dagger}\beta^{b}
\end{split}
\end{equation}
where $(i,a,b)$ are summed over.

The ground state energy of the Hamiltonian is $N\omega$. The ground state wave function is annihilated by all the annihilation operators. The unnormalized wave function can be written as,
\begin{equation}\label{groundstate}
\Psi = e^{-\omega \phi^{\dagger}\phi} e^{-\hbar TrQ^{2}} = e^{-\omega \phi^{\dagger}\phi} e^{ -\frac{TrA_{2}^{2}}{\hbar}}
\end{equation}
We can see that the width of the wave function along the gauge-field directions are of order, $\sqrt\hbar \sim \frac{1}{\sqrt k}$, and so are very small in the weak coupling or large $k$ limit.

\section{Singlet Sector Of The Hamiltonian}

We are interested in the states of the Hilbert space which are invariant under the global $SU(N)$ transformations of the quantum mechanical model. The singlet states in the Hilbert space can be obtained by acting on the ground-state with singlet creation operators. The "single-trace " operators are given by,
\begin{equation}
Tr\beta^{\dagger}\beta^{\dagger}, Tr\beta^{\dagger}\beta^{\dagger}\beta^{\dagger},...............,Tr(\beta^{\dagger})^{N}, \alpha^{\dagger}\bar\alpha^{\dagger}, \alpha^{\dagger}\beta^{\dagger}\bar\alpha^{\dagger},........., \alpha^{\dagger}(\beta^{\dagger})^{N}\bar\alpha^{\dagger}
\end{equation}
where we have defined the matrix $\beta^{\dagger} = \beta^{a\dagger}T^{a}$, which transforms in the adjoint representation and $\alpha^{\dagger}$ and $\bar\alpha^{\dagger}$ are column vectors transforming in the anti-fundamental and fundamental representations of the global symmetry group $SU(N)$.  Any other singlet creation operator can be written as linear combinations of products of these basis set of operators. The singlet states can be created by acting with the singlet operators on the ground state. In our case the ground state is exactly given by (\ref{groundstate}), and it is annihilated by all the annihilation operators. Now every state created by acting on the ground state with the singlet creation operators are not exact eigenfunctions of the total Hamiltonian\footnote{Our analysis will reveal that these states tend to be exact eigenstates in the large-N limit.}. States where no gauge field excitations are present are exact eigenfunctions of the total Hamiltonian, whereas the states with gauge field excitations are not in general exact eigenfunctions of the total Hamiltonian. They are exact eigenfunctions of the part of the Hamiltonian which contains no interaction term between scalar and gauge fields. This motivates us to separate the Hamiltonian into an interacting and a non-interacting part and treat the interacting part as perturbation. At this stage this separation is somewhat artificial because there is as such no small parameter in the Hamiltonian, \footnote{Although we are working in the large $k$ limit, $\frac{1}{k}$ is not a small parameter in the effective quantum mechanics problem. It is easy to show by properly scaling different variables that $\frac{1}{\sqrt k}$ is an overall multiplicative factor in the Hamiltonian.} but we shall justify this separation in the later part of this section. 

We define,
\begin{equation}
H_{0} = \omega(\alpha_{i}^{\dagger}\alpha_{i} + \bar\alpha_{i}^{\dagger}\bar\alpha_{i}) + \frac{\hbar}{2\omega} \beta^{a\dagger}\beta^{a} + N\omega
\end{equation}
and 
\begin{equation}
V = -\frac{\hbar}{2\omega}\biggl [\bar\alpha_{i}\lbrace T^{a},T^{b}\rbrace_{ij}\alpha_{j} + \alpha_{i}^{\dagger}\lbrace T^{a},T^{b}\rbrace_{ij}\bar\alpha_{j}^{\dagger}\biggr] \beta^{a\dagger}\beta^{b} = -\frac{\hbar}{2\omega} \tilde V
\end{equation}
\begin{equation}
V_{1} = \frac{\hbar}{2\omega} \biggl[\bar\alpha_{j}^{\dagger}\lbrace T^{a},T^{b}\rbrace_{ij} \bar\alpha_{i} + \alpha_{i}^{\dagger}\lbrace T^{a},T^{b}\rbrace_{ij} \alpha_{j}\biggr] \beta^{a\dagger}\beta^{b}
\end{equation}
The total Hamiltonian $H$ can be written as,
\begin{equation}
H = H_{0} + V + V_{1}
\end{equation}
It is easy to check that, $[H_{0},V_{1}] = 0$.

From the unperturbed Hamiltonian $H_{0}$ we can already see the basic dynamics of the system.  The zero point energy of the gauge fields creates a large frequency for the scalars $\omega^2 \sim N/k = \lambda$.  This is $N$ times larger than in the U(1) system.  This makes the scalar excitations heavy, with mass $\sim \omega$.  This large $\omega^2$ makes the size of $\phi^{\dagger} \phi$ smaller than in the U(1) case, $\phi^{\dagger} \phi\sim 1/\omega \sim \sqrt{k/N}$.  This is smaller by a factor of $\sqrt{N}$ than in the U(1) case.   The energy of the gauge field excitations is decreased by this factor.  The energy of such excitations is $\sim \sqrt{k/N}(1/k)  \sim \sqrt{\lambda}/N$.   This is the characteristic gap in the system.   In the large $N$ limit where we keep $\lambda$ fixed and let $N\rightarrow\infty$, the gauge field excitations are very light compared to the scalar field excitations.   We will show that the perturbative effects of the heavy scalars do not change these results.

More precisely, in the unperturbed theory the state with the smallest excitation energy can be written as,
\begin{equation}
|L> = Tr\beta^{\dagger}\beta^{\dagger}|\Omega> 
\end{equation}
where $|\Omega>$ is the ground state of  the system. The ground state is exact for the system whose wave function is given by (\ref{groundstate}). The state $|L>$ is non-degenerate and its energy is given by $\frac{\hbar}{\omega}\sim \frac{\sqrt\lambda}{N}$.   

We shall now study the effect of the terms $V_{1}$ and $V$ on the energy of this state and see that the parametric size of the gap is not changed.  

It is easy to see that the state $|L>$ is an exact eigenfunction of the Hamiltonian, $H_{0} + V_{1}$ with the same energy $\frac{\hbar}{\omega}$. 
The state is annihilated by $V_{1}$. Since $V_{1}$ commutes with the unperturbed Hamiltonian $H_{0}$ the effect of this term can be taken into account by diagonalizing $V_{1}$ restricted to degenerate eigenspaces of the unperturbed Hamiltonian $H_{0}$. Now $V_{1}$ annihilates any state which does not contain any scalar field excitation or gauge field excitations and so the energy and degeneracy of states containing only gauge field excitations or only scalar field excitations remain unchanged due to this term. We shall study the effect of this term later in this section. 

\subsection{Computation Of The Perturbation}
It is easy to check that the first order perturbation is zero. Before we write the results of the perturbation calculation we shall collect few results which are useful for our purpose. 

One can check the following results,
\begin{equation}\label{VL1}
\tilde V|L> =  \alpha^{\dagger}\beta^{\dagger}\beta^{\dagger}\bar\alpha^{\dagger} |\Omega> = |L1>
\end{equation}
\begin{equation}\label{VL2}
\tilde V|L1> = (\alpha^{\dagger}\beta^{\dagger}\bar\alpha^{\dagger})(\alpha^{\dagger}\beta^{\dagger}\bar\alpha^{\dagger}) |\Omega> + (\alpha^{\dagger}\beta^{\dagger}\beta^{\dagger}\bar\alpha^{\dagger})( \alpha^{\dagger}\bar\alpha^{\dagger}) |\Omega> + N (Tr\beta^{\dagger}\beta^{\dagger})|\Omega> 
\end{equation}
\begin{equation}\label{VL3}
\tilde V (\alpha^{\dagger}\beta^{\dagger}\bar\alpha^{\dagger})(\alpha^{\dagger}\beta^{\dagger}\bar\alpha^{\dagger}) |\Omega> = 2\alpha^{\dagger}\bar\alpha^{\dagger} (\alpha^{\dagger}\beta^{\dagger}\bar\alpha^{\dagger})(\alpha^{\dagger}\beta^{\dagger}\bar\alpha^{\dagger})|\Omega> + 2(N+1) \alpha^{\dagger}\beta^{\dagger}\beta^{\dagger}\bar\alpha^{\dagger} |\Omega>
\end{equation}
\begin{equation}\label{VL4}
\begin{split}
\tilde V (\alpha^{\dagger}\beta^{\dagger}\beta^{\dagger}\bar\alpha^{\dagger})( \alpha^{\dagger}\bar\alpha^{\dagger}) |\Omega>
& = \alpha^{\dagger}\bar\alpha^{\dagger} (\alpha^{\dagger}\beta^{\dagger}\bar\alpha^{\dagger})^{2}|\Omega> + (\alpha^{\dagger}\bar\alpha^{\dagger})^{2} (\alpha^{\dagger}\beta^{\dagger}\beta^{\dagger}\bar\alpha^{\dagger})|\Omega>  \\
 &  + (N+2) (\alpha^{\dagger}\beta^{\dagger}\beta^{\dagger}\bar\alpha^{\dagger})|\Omega>  + (N+1) \alpha^{\dagger}\bar\alpha^{\dagger} Tr\beta^{\dagger}\beta^{\dagger} |\Omega>
 \end{split}
\end{equation}
In deriving these results we have made use of the following relations,
\begin{equation}
\sum_{a=1}^{N^{2}-1}(T^{a})_{ij}(T^{a})_{kl} = \frac{1}{2} (\delta_{il}\delta_{jk} - \frac{1}{N} \delta_{ij}\delta_{kl})
\end{equation}
and 
\begin{equation}
Tr(T^{a}T^{b}) = \frac{1}{2}\delta^{ab}
\end{equation}
In (\ref{VL3}) one can neglect the $\frac{1}{N}$ piece in the large $N$ limit. The generators are taken in the fundamental representation.

The states appearing in eqns (\ref{VL1})-(\ref{VL4}) are unnormalized. The norms of these states are given by,
\begin{equation}\label{norm1}
||Tr\beta^{\dagger}\beta^{\dagger}|\Omega>||^{2}\sim N^{2}
\end{equation}
\begin{equation}\label{norm2}
||\alpha^{\dagger}\beta^{\dagger}\beta^{\dagger}\bar\alpha^{\dagger} |\Omega>||^{2} \sim N^{3}
\end{equation}
\begin{equation}
||(\alpha^{\dagger}\beta^{\dagger}\bar\alpha^{\dagger})(\alpha^{\dagger}\beta^{\dagger}\bar\alpha^{\dagger}) |\Omega>||^{2} \sim N^{4}
\end{equation}
\begin{equation}
||(\alpha^{\dagger}\beta^{\dagger}\beta^{\dagger}\bar\alpha^{\dagger})( \alpha^{\dagger}\bar\alpha^{\dagger}) |\Omega> ||^{2} \sim N^{4}
\end{equation}
\begin{equation}
||(\alpha^{\dagger}\bar\alpha^{\dagger})^{2} (\alpha^{\dagger}\beta^{\dagger}\beta^{\dagger}\bar\alpha^{\dagger})|\Omega> ||^{2} \sim N^{5}
\end{equation}
\begin{equation}
|| \alpha^{\dagger}\bar\alpha^{\dagger} (\alpha^{\dagger}\beta^{\dagger}\bar\alpha^{\dagger})^{2}|\Omega>|| ^{2} \sim N^{5}
\end{equation}
The inner product of states containing different numbers of scalar and gauge excitations are orthogonal. The inner product of the states appearing in (\ref{norm1}) and (\ref{norm2}) are given by,
\begin{equation}
\biggl((\alpha^{\dagger}\beta^{\dagger}\bar\alpha^{\dagger})(\alpha^{\dagger}\beta^{\dagger}\bar\alpha^{\dagger}) |\Omega> ,(\alpha^{\dagger}\beta^{\dagger}\beta^{\dagger}\bar\alpha^{\dagger})( \alpha^{\dagger}\bar\alpha^{\dagger}) |\Omega> \biggr) \sim N^{3}
\end{equation}
\begin{equation}
\biggl( (\alpha^{\dagger}\bar\alpha^{\dagger})^{2} (\alpha^{\dagger}\beta^{\dagger}\beta^{\dagger}\bar\alpha^{\dagger})|\Omega> ,  \alpha^{\dagger}\bar\alpha^{\dagger} (\alpha^{\dagger}\beta^{\dagger}\bar\alpha^{\dagger})^{2}|\Omega>\biggl) \sim N^{4}
\end{equation}
So these two states are orthogonal in the large-$N$ limit. In particular, the normalized states\footnote{We shall write down only powers of $N$ that appear in the normalization of the states in the large-$N$ limit. There are $O(1)$ numbers which multiply the states in the large-$N$ limit. We shall not write them because they are not important for our purpose, at least to the order we are working.} (in the large-$N$ limit) containing two gauge field excitations are of the form,
\begin{equation}\label{normal1}
\frac{1}{N}Tr(\beta^{\dagger}\beta^{\dagger})|\Omega>
\end{equation}
\begin{equation}\label{normal2}
\frac{1}{N^{\frac{3}{2}}}\alpha^{\dagger}\beta^{\dagger}\beta^{\dagger}\bar\alpha^{\dagger} |\Omega>
\end{equation}
\begin{equation}\label{normal3}
\frac{1}{N^{\frac{3}{2}}} (\alpha^{\dagger}\bar\alpha^{\dagger}) Tr\beta^{\dagger}\beta^{\dagger}|\Omega>
\end{equation}
\begin{equation}\label{normal4}
\frac{1}{N^{2}}(\alpha^{\dagger}\beta^{\dagger}\bar\alpha^{\dagger})(\alpha^{\dagger}\beta^{\dagger}\bar\alpha^{\dagger}) |\Omega>
\end{equation}
\begin{equation}\label{normal5}
\frac{1}{N^{2}}(\alpha^{\dagger}\beta^{\dagger}\beta^{\dagger}\bar\alpha^{\dagger})( \alpha^{\dagger}\bar\alpha^{\dagger}) |\Omega>
\end{equation}
\begin{equation}\label{normal6}
\frac{1}{N^{2}} (\alpha^{\dagger}\bar\alpha^{\dagger})(\alpha^{\dagger}\bar\alpha^{\dagger}) (Tr\beta^{\dagger}\beta^{\dagger})|\Omega>
\end{equation}
\begin{equation}\label{normal7}
\frac{1}{N^{\frac{5}{2}}} \alpha^{\dagger}\bar\alpha^{\dagger} (\alpha^{\dagger}\beta^{\dagger}\bar\alpha^{\dagger})^{2}|\Omega>
\end{equation}
\begin{equation}\label{normal8}
\frac{1}{N^{\frac{5}{2}}} (\alpha^{\dagger}\bar\alpha^{\dagger})^{2} (\alpha^{\dagger}\beta^{\dagger}\beta^{\dagger}\bar\alpha^{\dagger})|\Omega>
\end{equation}
There are an infinite number of such states. These states are all normalized and mutually orthogonal in the large-$N$ limit. This is true even in the interacting theory because the annihilation operators are defined with respect to the exact ground state of the interacting theory. These states are also exact eigenfunctions of the unperturbed Hamiltonian $H_{0}$. 
\subsection{1-st Order Perturbation}
It is easy to see that the the first order perturbation is zero.
\subsection{2-nd Order Perturbation}
The answer for the second order perturbation is,
\begin{equation}\label{shift}
\Delta^{(2)} = -\frac{1}{2}\frac{\hbar}{\omega}
\end{equation}
where $\Delta^{(2)}$ is the second order shift in the energy of the state $Tr\beta^{\dagger}\beta^{\dagger}|\Omega>$. We can see that the second order shift is of the same order of magnitude as the zeroth order energy of the state which is $\frac{\hbar}{\omega}$.
\subsection{3-rd Order Perturbation}
The formula for the third order energy shift is,
\begin{equation}
\Delta^{(3)} =  \sum_{k\neq n, m\neq n} \frac{V_{nk}V_{km}V_{mn}}{E_{nk}E_{nm}} - V_{nn} \sum_{k\neq n}\frac{|V_{nk}|^{2}}{E_{nk}^{2}}
\end{equation}
where, $V_{nk}= <n|V|k>$ and $E_{mn} = E_{m} - E_{n}$. The states and energies are all referred to the unperturbed Hamiltonian $H_{0}$ and all the states are normalized.

In our case, $|n> = \frac{1}{N}|L> = \frac{1}{N}Tr\beta^{\dagger}\beta^{\dagger}|\Omega>$ and so $V_{nn} = 0$. Now from eqn-(6.7) we get,
\begin{equation}
V|n> = -\frac{1}{N} \frac{\hbar}{2\omega} \alpha^{\dagger}\beta^{\dagger}\beta^{\dagger}\bar\alpha^{\dagger} |\Omega> = \frac{1}{N} |L1>
\end{equation}
Now $|L1>$ is an exact eigenfunction of $H_{0}$ and so it follows from our previous discussion that the matrix element $V_{nk}$ and $V_{mn}$ are nonzero only if $|k> = |m> \propto |L1>$. But in that case the matrix element $V_{km} =0$. So the third order perturbation vanishes.
\subsection{4-th Order Perturbation}
The formula for the fourth order perturbation is,
\begin{equation}
\Delta^{(4)} = \sum_{k_{2}\neq n, k_{3}\neq n, k_{4}\neq n}\frac{V_{nk_{2}}V_{k_{2}k_{3}}V_{k_{3}k_{4}}V_{k_{4}n}}{E_{nk_{2}}E_{nk_{3}}E_{nk_{4}}} - \sum_{k_{1}\neq n, k_{2}\neq n}\frac{V_{nk_{1}}V_{k_{1}n}}{E_{nk_{1}}^{2}} \frac{V_{nk_{2}}V_{k_{2}n}}{E_{nk_{2}}}
\end{equation}
We have not written down the terms which are multiplied by $V_{nn}$ which is zero in our case. Let us study the contribution of the first term . Using the same argument as in the case of the third order perturbation, we conclude that $|k_{2}> = |k_{4}> \propto |L1>$. if this condition is not satisfied the first term will vanish. Using this the first term can be simplified to,
\begin{equation}
\sum_{k_{3}\neq n}\frac{|V_{nL1}|^{2}|V_{L1k_{3}}|^{2}}{E_{nL1}E_{nk_{3}}E_{nL1}} 
\end{equation}
Now we shall calculate the $N$ scaling of these matrix elements in the large $N$ limit.
\begin{equation}
\begin{split}
V_{L1n} \sim <\alpha^{\dagger}\beta^{\dagger}\beta^{\dagger}\bar\alpha^{\dagger}|\frac{1}{N^{\frac{3}{2}}} V \frac{1}{N} |Tr\beta^{\dagger}\beta^{\dagger}> & =  -\frac{\hbar}{2\omega}\frac{1}{N^{\frac{5}{2}}} <\alpha^{\dagger}\beta^{\dagger}\beta^{\dagger}\bar\alpha^{\dagger}|\alpha^{\dagger}\beta^{\dagger}\beta^{\dagger}\bar\alpha^{\dagger}>  \\ & \sim -\frac{1}{\sqrt{Nk}} \frac{1}{N^{\frac{5}{2}}} (N^{3} + O(N^{2}))\\& = -\frac{1}{\sqrt k}(1 + O(\frac{1}{N}))
\end{split}
\end{equation}
Now we have to compute the matrix element $V_{k_{3}L1}$. From (\ref{VL2})we get,
\begin{equation}
V|L1> = \biggl(\frac{\hbar}{2\omega}\biggr)^{2} \biggl((\alpha^{\dagger}\beta^{\dagger}\bar\alpha^{\dagger})(\alpha^{\dagger}\beta^{\dagger}\bar\alpha^{\dagger}) |\Omega> + (\alpha^{\dagger}\beta^{\dagger}\beta^{\dagger}\bar\alpha^{\dagger})( \alpha^{\dagger}\bar\alpha^{\dagger}) |\Omega> + N (Tr\beta^{\dagger}\beta^{\dagger})|\Omega> \biggr)
 \end{equation}
Now every state appearing in the above equation is an exact eigenstate of $H_{0}$ and they are mutually orthogonal at least in the large-N limit. So the matrix element can be nonzero only if the the state $|k_{3}>$ is one of the three states appearing in the formula. Now $|k_{3}>$ cannot be the last state because it is proportional to the state $|n>$. So $|k_{3}>$ can be any one of the remaining two states. Let us first take,
\begin{equation}
|k_{3}> \propto (\alpha^{\dagger}\beta^{\dagger}\bar\alpha^{\dagger})(\alpha^{\dagger}\beta^{\dagger}\bar\alpha^{\dagger}) |\Omega>
\end{equation} 
So,
\begin{equation}
\begin{split}
V_{k_{3}L1} &\sim <(\alpha^{\dagger}\beta^{\dagger}\bar\alpha^{\dagger})(\alpha^{\dagger}\beta^{\dagger}\bar\alpha^{\dagger})|\frac{1}{N^{2}} V \frac{1}{N^{\frac{3}{2}}} |\alpha^{\dagger}\beta^{\dagger}\beta^{\dagger}\bar\alpha^{\dagger}>   \\ &  \sim - \frac{\hbar}{2\omega} \frac{1}{N^{\frac{7}{2}}} <(\alpha^{\dagger}\beta^{\dagger}\bar\alpha^{\dagger})(\alpha^{\dagger}\beta^{\dagger}\bar\alpha^{\dagger})|(\alpha^{\dagger}\beta^{\dagger}\bar\alpha^{\dagger})(\alpha^{\dagger}\beta^{\dagger}\bar\alpha^{\dagger})>  \\ & \sim - \frac{1}{\sqrt{Nk}}\frac{1}{N^{\frac{7}{2}}} (N^{4} + O(N^{3})) \sim -\frac{1}{\sqrt k} (1 + O(\frac{1}{N}))
\end{split}
\end{equation}
The same scaling holds for the other choice of $|k_{3}>$. So we can conclude that the matrix elements scale like $-\frac{1}{\sqrt k}(1+ O(\frac{1}{N}))$ and so the leading contribution in the large N limit is $-\frac{1}{\sqrt k}$.

The energy denominators are all of order $\sqrt\omega \sim \sqrt \lambda = \sqrt{\frac{N}{k}}$, because the states appearing in the formula other than $|n>$ contain scalar excitations. So the leading contribution of the matrix element in the large-$N$ limit is of order,
\begin{equation}
(-\frac{1}{\sqrt k})^{4} \frac{1}{(\sqrt\lambda)^{3}} =  \frac{1}{k^{2}} \frac{1}{\frac{N^{\frac{3}{2}}}{k^{\frac{3}{2}}}} = \frac{1}{\sqrt{Nk}}\frac{1}{N}
\end{equation}
So we can see that this contribution is $\frac{1}{N}$ suppressed compared to the second order contribution. It is easy to see that the second term in formula (\ref{normal5}) gives the same $\frac{1}{N}$ suppressed contribution.
\subsection{5-th Order Perturbation}
It is easy to convince oneself that the 5-th order perturbation also vanishes for the same reason that the third order perturbation vanished. The fifth order perturbation contains terms of two kinds. One kind of terms is multiplied by $V_{nn}$ which is identically zero in our case. The second kind of terms are all multiplied by the matrix element appearing in the third order perturbation\footnote{The formalism of time-independent perturbation theory can be used to determine the 4-th, 5-th and 6-th order perturbations. References on this formalism are presented in the classic texts\cite{LL3:1977,Sakurai:1167961}. The results for higher order perturbations are stated in \cite{Wheeler_2000}.} and so is identically zero in our case. The only term which survives is the following,
\begin{equation}
V_{nk_{1}}V_{k_{1}k_{2}}V_{k_{2}k_{3}}V_{k_{3}k_{4}}V_{k_{4}n}
\end{equation}
The energy denominator is also there and the indices except $n$ is summed over subject to the same constraint. So by following the same argument as in the previous case we conclude that $|k_{1}> = |k_{4}> \propto |L1>$. So the states $|k_{2}>$ and $|k_{3}>$ must belong to the subspace spanned by the states $(\alpha^{\dagger}\beta^{\dagger}\bar\alpha^{\dagger})(\alpha^{\dagger}\beta^{\dagger}\bar\alpha^{\dagger}) |\Omega> $ and $(\alpha^{\dagger}\beta^{\dagger}\beta^{\dagger}\bar\alpha^{\dagger})( \alpha^{\dagger}\bar\alpha^{\dagger}) |\Omega>$. These states are orthogonal in the large N limit. Now the matrix element $V_{k_{2}k_{3}}$ vanishes in this subspace. So the fifth order perturbation is identically zero.
\subsection{6-th Order Perturbation}
In this case one can show using the results stated in the previous sections that the contribution goes like,
\begin{equation}
\frac{1}{\sqrt {kN}} \frac{1}{N^{2}} , \\\\ N\rightarrow\infty
\end{equation}
\subsection{The Odd Order Perturbation Is Zero To All Orders}
Let us consider the $(2p+1)$-th order perturbation theory. The $(2p+1)$-th order perturbation contains the term,
\begin{equation}
V_{nk_{1}}V_{k_{1}k_{2}}\ldots V_{k_{p-1}k_{p}}V_{k_{p}k_{p+1}}V_{k_{p+1}k_{p+2}}...............V_{k_{2p}n}
\end{equation}
There is an energy denominator and the intermediate states $|k_{i}>$ do not take the value $|n> = Tr\beta^{\dagger}\beta^{\dagger}|\Omega>$. Let us denote by $A$ and $B$ the following matrix elements,
\begin{equation}
A = V_{nk_{1}}V_{k_{1}k_{2}}\ldots V_{k_{p-1}k_{p}}
\end{equation}
and 
\begin{equation}
B = V_{k_{p+1}k_{p+2}}\ldots V_{k_{2p}n}
\end{equation}
Since $V$ is Hermitian, the complex conjugate of $B$ can be written as,
\begin{equation}
B^{*} = V_{nk_{2p}}\ldots V_{k_{p+2}k_{p+1}}
\end{equation}
Both of these matrix elements represent the following process. The potential $V$ creates or annihilates two scalar excitations, one of type $\alpha$ and another of type $\bar\alpha$. Since we are considering only singlet states,\footnote {We are starting with the singlet state $|n>$ and and since $V$ is a singlet operator we never leave the singlet sector.} every state contains an equal number of $\alpha$ and $\bar\alpha$ excitations and so the total number of scalar excitations is always an even integer. The action of $V$ increases or decreases this integer in steps of $2$. The number of gauge excitations does not change because because $V$ contains a creation and an annihilation operator for the gauge excitations. More precisely the number operator for the gauge oscillators given by $Tr\beta^{\dagger}\beta$ commutes with $V$. Since both $A$ and $B^{*}$ represent the same physical process let us concentrate on $A$. So the systems starts at the state $|k_{p}>$ with some number of scalar excitations, say $2m$, and after a series of transitions it ends up in the state $|n>$ with zero scalar excitations. If the end state has nonzero scalar excitations then the first matrix element $V_{nk_{1}}$ vanishes and this term in the perturbation series is zero. The system can make a total of $p$ transitions and some of them are up transitions and some of them are down transitions where the number of scalar field quanta increases or decreases by 2, respectively. Let $n_{+}$ and $n_{-}$ be the number of up and down transitions. So the they have to obey the following relations,
\begin{equation}
n_{+} + n_{-} = p
\end{equation}
 and 
\begin{equation}
2(n_{+} - n_{-}) = 2m
\end{equation}
The solution is given by,
\begin{equation}
n_{+} = \frac{p+m}{2}, \\  n_{-} = \frac{p-m}{2}
\end{equation}
 Now $p$ is a fixed integer at a given order and so what can vary is the integer $m$ which determines the number the scalar excitations in the state $|k_{p}>$. Let $m=m_{0}$ be a value for which the matrix element $A$ is nonzero, i.e, the system can make $p$ transitions to reach a state with no scalar excitations. Now $n_{+}$ and $n_{-}$ are integers. So the next nearest values of $m$ for which the matrix element  $A$ is nonzero is given by $m_{0}\pm 2$. It is not $m_{0}\pm 1$ because in that case $n_{\pm}$ will be half-integers. So if $|k_{p}> = |2m_{0}>$ is one state then the nearest states are $|k_{p}'> = |2m_{0}>$ or $|k_{p}'> = |2m_{0}\pm 4>$. the same argument goes through for the amplitude $B^{*}$. Now we have the matrix element $V_{k_{p}k_{p+1}}$. This matrix element will be nonzero only if the states $|k_{p}>$ and $|k_{p+1}>$ differ by two units of scalar excitations. So $|k_{p+1}>$ has to be a state of the form $|2m_{0}\pm 2>$. But in that case we know that the matrix element $B^{*}$ will vanish, because if $|2m_{0}>$ is a valid state then the next nearest states are $|2m_{0}\pm 4>$. So in any case the total matrix element has to be zero. So the odd order perturbation contribution is zero to all orders.
 
\subsection{Gap In The System}
In the large-$N$ limit the energy of the state $Tr\beta^{\dagger}\beta^{\dagger}|\Omega>$ can be written as,
\begin{equation}
\Delta = \frac{\hbar}{\omega} \biggl( 1 + 0 - \frac{1}{2} + 0 + \frac{a_{4}}{N} + 0 + \frac{a_{6}}{N^{2}} + 0 +................\biggr)
\end{equation} 
where $a_{4}$ and $a_{6}$ are $O(1)$ numbers. This expression justifies our treatment of the potential $V$ as perturbation in the large-$N$ limit. So in the large-$N$ limit the leading term in the gap is 
\begin{equation}
\Delta = \frac{\hbar}{2\omega} \sim \frac{\sqrt\lambda}{N} 
\end{equation}

\subsection{Effect Of The Perturbation \texorpdfstring{$V_{1}$}{V1}}
The potential $V_{1}$ is a gauge singlet and it commutes with the Hamiltonian $H_{0}$. Now instead of treating $H_{0}$ as the unperturbed Hamiltonian we could have treated $H_{0} + V_{1}$ as the unperturbed Hamiltonian. It is easy to see that the exact ground state of the total Hamiltonian is also an exact eigenstate of the Hamiltonian $H_{0} + V_{1}$ with the same eigenvalue because the ground state $|\Omega>$ is annihilated by the potential $V_{1}$. In fact only states which contain both the scalar and gauge excitations are not annihilated by the potential $V_{1}$. So states containing either scalar or gauge excitations only, have the same energy when thought of as eigenstates of the Hamiltonian $H_{0}+V_{1}$. To proceed we need to consider states which contain both scalar and gauge excitations. Since $H_{0}$ and $V_{1}$ commute, to compute the change in the energy we just need to diagonalize the potential $V_{1}$ in a given eigenspace of the Hamiltonian $H_{0}$. 

	The singlet sector of $H_0$ eigenvectors is degenerate as can be seen in eqns.(\ref{normal2}) -(\ref{normal8}). We must therefore construct specific linear combinations of these $H_0$ eigenvectors to simultaneously diagonalize $V_1$. This is necessary since the naive singlet sector harmonic oscillator basis does not diagonalize $V_1$ but only reduces the operator to a block diagonal form, with the blocks corresponding to degenerate eigenspaces of $H_{0}$.

For the singlet sector in the large $N$ limit, the eigenvectors of $V_1$ that correspond to a particular block are composed of eigenvectors of $H_0$ with the same $N$ scaling of their norm. For example (\ref{normal3}) and (\ref{normal4}) both have normalizations $N^{-3/2}$ due to the fact that their inner product sans normalization goes as $N^3$ in the large $N$ limit. This implies that the block that contains these states is a $2$-dim subspace. We shall label eigenstates of this space, which are composed of linear combinations of $\alpha^{\dagger}\beta^{\dagger}\beta^{\dagger}\bar{\alpha}^{\dagger}|\Omega>$ and $\alpha^{\dagger}\bar{\alpha}^{\dagger}Tr(\beta^{\dagger}\beta^{\dagger})|\Omega>$ as $|N^3_{(i)}>$. The $N^3$-norm states are up to a normalization,

\begin{align}
|N^{3}_{(1)}> & \sim\Big [\alpha^{\dagger}\beta^{\dagger}\beta^{\dagger}\bar{\alpha}^{\dagger} \Big]|\Omega>\label{v11}\\
|N^{3}_{(2)}>&\sim\Big [\alpha^{\dagger}\beta^{\dagger}\beta^{\dagger}\bar{\alpha}^{\dagger} - N\alpha^{\dagger}\bar{\alpha}^{\dagger}Tr(\beta^{\dagger}\beta^{\dagger})\Big]|\Omega>\label{v12}
\end{align}

with eigenvalues $\frac{\hbar}{2\omega}N$ and $0$ respectively\footnote{ We are expressing eigenvalues and eigenvectors in the large $N$ limit which means the expressions are ignoring any additive terms of sub-leading order in $N$. For example the eigenvalues resulting in (\ref{v11}) and (\ref{v12}) have lower order contributions besides what is shown.}. Recalling that $\frac{\hbar}{2\omega} = \sqrt{\frac{32\pi}{ Nk}}$ we see in large $N$ the first eigenvalue goes as $\sqrt{\frac{\pi N}{2 k}}$ while the other is approximately zero.

Similarly the states with unnormalized inner products scaling as $N^4$, form a $3\times3$ block. At the $N^{4}$ level in the large $N$ limit we find a zero eigenvalue. The two nonzero eigenvalues of $V_1$ at this level being
\begin{equation}
\sqrt{\frac{8\pi}{ k}}\sqrt{4N}\hspace{1in}\frac{1}{2}\sqrt{\frac{32\pi}{\alpha k}}\frac{\sqrt{N}}{2}.
\end{equation}

These correspond to the eigenvectors
\begin{align}
|N^{4}_{(1)}> & \sim\Big [-2N^{3/2}\alpha^{\dagger}\beta^{\dagger}\bar{\alpha}^{\dagger}\alpha^{\dagger}\beta^{\dagger}\bar{\alpha}^{\dagger} +2N\alpha^{\dagger}\bar{\alpha}^{\dagger}\alpha^{\dagger}\beta^{\dagger}\beta^{\dagger}\bar{\alpha}^{\dagger} + \alpha^{\dagger}\bar{\alpha}^{\dagger}\alpha^{\dagger}\bar{\alpha}^{\dagger}Tr(\beta^{\dagger}\beta^{\dagger})\Big]|\Omega>\\
|N^{4}_{(2)}>&\sim\Big [-2N^{1/2}\alpha^{\dagger}\beta^{\dagger}\bar{\alpha}^{\dagger}\alpha^{\dagger}\beta^{\dagger}\bar{\alpha}^{\dagger} +N\alpha^{\dagger}\bar{\alpha}^{\dagger}\alpha^{\dagger}\beta^{\dagger}\beta^{\dagger}\bar{\alpha}^{\dagger} + \alpha^{\dagger}\bar{\alpha}^{\dagger}\alpha^{\dagger}\bar{\alpha}^{\dagger}Tr(\beta^{\dagger}\beta^{\dagger})\Big]|\Omega>
\end{align}

up to a normalization.

 To obtain the eigenvector corresponding to the zero eigenvalue, one should find the appropriate linear combination of $\alpha^{\dagger}\beta^{\dagger}\bar{\alpha}^{\dagger}\alpha^{\dagger}\beta^{\dagger}\bar{\alpha}^{\dagger}|\Omega>$, $\alpha^{\dagger}\bar{\alpha}^{\dagger}\alpha^{\dagger}\beta^{\dagger}\beta^{\dagger}\bar{\alpha}^{\dagger}|\Omega>$, and $\alpha^{\dagger}\bar{\alpha}^{\dagger}\alpha^{\dagger}\bar{\alpha}^{\dagger}Tr(\beta^{\dagger}\beta^{\dagger})|\Omega>$ that is orthogonal to $|N^{4}_{(1)}>$ and $|N^{4}_{(2)}>$.

For the $N^{5}$ level, there is also a $3\times3$ block of singlet states. This time we only have one non-zero eigenvalue in the large $N$ limit
\begin{equation}
\frac{3}{2}\sqrt{\frac{2\pi}{ k}}\sqrt{N}
\end{equation}

corresponding to the eigenvector,

\begin{equation}
|N^{5}> \sim\Big[\frac{3N^{3/2}}{2} \alpha^{\dagger}\Bar{\alpha}^{\dagger}\alpha^{\dagger}\beta^{\dagger}\bar{\alpha}^{\dagger}\alpha^{\dagger}\beta^{\dagger}\bar{\alpha}^{\dagger} + \alpha^{\dagger}\Bar{\alpha}^{\dagger}\alpha^{\dagger}\Bar{\alpha}^{\dagger}\alpha^{\dagger}\Bar{\alpha}^{\dagger}\alpha^{\dagger}Tr(\beta^{\dagger}\beta^{\dagger})\Big]|\Omega>.
\end{equation}
The zero eigenvalues correspond to vectors spanning the plane orthogonal to $|N^{5}>$ in the $\alpha^{\dagger}\Bar{\alpha}^{\dagger}\alpha^{\dagger}\beta^{\dagger}\bar{\alpha}^{\dagger}\alpha^{\dagger}\beta^{\dagger}\bar{\alpha}^{\dagger}|\Omega>$,  $\alpha^{\dagger}\Bar{\alpha}^{\dagger}\alpha^{\dagger}\Bar{\alpha}^{\dagger}\alpha^{\dagger}\Bar{\alpha}^{\dagger}\alpha^{\dagger}Tr(\beta^{\dagger}\beta^{\dagger})|\Omega>$, $\alpha^{\dagger}\bar{\alpha}^{\dagger}\alpha^{\dagger}\bar{\alpha}^{\dagger}\alpha^{\dagger}\beta^{\dagger}\beta^{\dagger}\bar{\alpha}^{\dagger}|\Omega>$ basis, up to a normalization.

We can see from the above analysis that the shifts in the energies of few low lying states is always positive semi-definite. We see that the $V_1$ eigenvalues in the singlet sector do not ruin the analysis of the previous section. The eigenvalues are positive semi-definite in the large-$N$ limit and make the perturbative analysis of $V$ more robust.

\subsection{Counting Of States}
Since in the large-N limit the states containing scalar excitations are much heavier than the states containing pure gauge excitations we can compute the number of states in the low energy sector by counting only the states where there is no scalar excitation.  (The effect in equation \ref{shift}  should be representable as a shift in the pure gauge harmonic oscillator frequency.)
This is exactly the counting problem for the quantum mechanics of one hermitian matrix model with harmonic potential which has been solved in the classic paper \cite{Brezin:1977sv}.  That model in the singlet sector is exactly equivalent to $N$ free fermions in a harmonic potential. We note that the Fermi energy of the system is given by, $E_F = \sqrt\lambda$. Again this problem is easy to solve.

More precisely, we can take a limit with $k\to\infty$ and scale the temperature
so that the low-energy states of the Chern-Simons/matter-zero-mode system, and
only those states, contribute in the thermal ensemble. If we take $k$ large and
 scale the temperature $T$ according to the limit
\begin{equation}
{E_F = \sqrt\lambda} \ll T \ll m_{\rm scalar}\text{ ,}
\end{equation}
then the Boltzmann factor suppressing the contributions of the KK modes is $\exp{\{- \frac{{m_{\rm scalar}}}{T}\}}$, so the KK modes can be ignored altogether; while the number of $\beta$-oscillator  states with energies below the temperature $\delta$ grows as $\propto ( {T  \sqrt{k} })^N / N!$ and the entropy is given by,

\begin{equation}\label{ourentropyb}
S \simeq N \big [  \log (T/\sqrt{\lambda}) + O(1) \big ].
 \end{equation} 
  At this temperature, a simple semiclassical argument suffices to derive this scaling.
 The harmonic oscillator frequency goes as $\omega \equiv 1 / \sqrt{Nk}$.
  The classical partition function
for $N$ identical harmonic oscillators goes as
\begin{equation}
Z_{\rm classical}
= {\frac{1}{N!}}\int dp\llo i dq\llo i \expp{- \sum\llo i (p\llo i \sqd + \omega\sqd q\llo i\sqd) / T}\ .
\end{equation}
Upon performing the integral, one gets a $T\uu 1$ for each $p,q$ pair of 
integrals, and one gets an $\omega\uu{-1}$ from 
every $q$ integral.  The total factor is then $(\pi T / \omega)\uu N / N!$.  The log of that is
$ {\rm ln}(Z_{\rm classical}) =
N ~ {\rm ln}(T / \sqrt{\lambda})$.  This gives (\ref{ourentropyb}) exactly.
We observe that this is well-behaved in the 't Hooft limit.

We can also consider a different limit where the temperature is much less than the Fermi energy. In this regime the entropy of the system is given by,
\begin{equation}
S \sim \frac{NT}{\sqrt \lambda}
\end{equation}  
This is the leading term in the Sommerfeld expansion of the entropy.


\section{Discussion}

\subsection{New Light States}

We have shown that on a spatial $T^2$  the matter-Chern-Simons conformal field
theory based on a single scalar field in the
fundamental representation has a set of low-lying states with energy gaps of order ${\frac{1}{{\sqrt{
Nk}}}}$ (for the free scalar) or $\frac{1}{k}$ (for the critical scalar).  As a result, there is a divergent degeneracy of states in the limit where
the level $k$ goes to infinity, at fixed rank $N$ of the gauge group.

The Vasiliev theory successfully describes correlation functions of
higher-spin conserved currents of the
infinite$-k$ limit on $\IR^3$, as well as its partition function on $S\uu 1 \times S\uu 2$. However a consistent proposal for a gravitational dual description for 
the Chern-Simons-matter CFT analyzed in this article
should provide a bulk realization for the CFT partition function on general boundary geometry, including the light states we have found and the parametrically large (in $k$) entropy associated with them.   The entropy of our system on
$T\uu 2$ is (\ref{OurEntropyA}), which diverges for fixed $N$ and
large $k$.  This agrees with the large-$k$ entropy of the pure Chern-Simons sector
(\cite{Witten:1999ds}, \cite{Elitzur:1989nr}) 
\begin{equation}
{\rm ln}(Z) \simeq   (N-1)~ {\rm ln}(k) - {\rm ln}((N-1)!) + O(k\uu{-1})\ .
\label{CSDegenTorus}\end{equation}
which for large $N$ is
\begin{equation}
{\rm ln}(Z) \propto N ~{\rm ln}(\lambda\uu{-1})\ .
\end{equation}
The addition of matter does not affect the parametric $N~{\rm ln}(k)$ divergence of
the entropy. 

\subsection{Vasiliev As A imit Of String Theory}

It is clear that the Vasiliev theory does not by itself contain the
degrees of freedom corresponding to the large entropy of the CFT on spatial
slices of genus $g =1$.   Nor can any deformation of the theory with deforming interaction 
terms in the action or equations of motion that are proportional to positive powers of $\lambda$
(for example \cite{YinSlides, GiombiSlides}).
 There are no fundamental fields of Vasiliev theory that
could generate such a topology-dependent divergence, and any solitonic collective
excitations should have masses that scale with negative, rather than positive powers of
$\lambda$.  The proposal to derive Vasiliev gravity as a limit of string theory in which
stringy physics decouples altogether, appears to work under certain circumstances, but
not universally.  The limit $\lambda \to 0$ is not a conventional decoupling limit for
string theory like the infinite tension limit $\alpha^\prime\to 0$, where string oscillator excitations  decouple in the usual Wilsonian sense.
Rather, $\lambda\to 0$ can be thought of as a limit in which the string tension goes
to zero and each string "bit" moves as an independent particle.  However in non-simply-connected
spaces, there is a topological constraint which does not allow all the string bits
to move independently, when the string winds a noncontractible cycle.  As a result,
there is an infinite tower of independent states distinguished by their winding, but with a parametrically low cost for states with arbitrarily large winding. 

Holographic duality suggests that the large-$k$ divergence is related to an
incompleteness of the Vasiliev theory.  For a bulk with
boundary $T\uu 3$, there is a singular solution of Einstein gravity with negative cosmological
constant that is also a solution of Vasiliev gravity, in which the spatial $T\uu 2$ shrinks
to zero size.  It is natural to associate this singularity to the light states\footnote{We thank Tom Banks for conversations on this point}.   Wrapped strings and T duality resolve this singularity in standard bulk string theory situations, and we infer that a consistent ultraviolet
completion of the Vasiliev action is likely to involve string degrees of freedom to
account for the entropy.
The connection between the large-$k$ degeneracy and nonvanishing
fundamental group, for instance, may suggest an identification of our light states with the closed
string sector of the topological open-closed string theory proposed in \cite{YinSlides, TopologicalStringVasiliev}.

\subsection{Higher Genus}
For higher genus, $g\geq 2$, the quartic
interaction in the Wilson-Fisher theory
stabilizes the scalars independent of $k$ against their conformal
coupling $\frac{1}{8}~{\tt Ricci}_3~\phi\sqd$ to the Ricci scalar
interaction, and gives their energies
a gap of order $1$.  The quantum mechanical techniques discussed above should
apply here and give gaps in the gauge field sector of order $\frac{1}{k}$.
As discussed earlier, for a massive scalar field we expect that at 
temperatures $  \exp(-M)  \ll T \ll 1$ the entropy
should reduce to that of the pure Chern-Simons system.   Here  the effective $M \sim 1$ so we do not have parametric control, but the largest effect the matter field could have is if its vev where large, effectively Higgsing the system down to an SU(N-1) pure Chern-Simons theory\footnote{This is what will happen in the free massless scalar system where the  effect of the $\frac{1}{8}~{\tt Ricci}_3~\phi\sqd$ term is stabilized by the $\lambda^2 \phi^6$  discussed in \cite{Aharony:2011jz}  at a large vev  of order $\phi^2 \sim \frac{1}{\lambda}$ . \ } .  So we can use the pure Chern-Simons entropy as a good estimate of the entropy of our system at large $N$.

The entropy of the Chern-Simons theory on surfaces of genus $g$
\cite{{VerlindeProceedings}}\footnote{as cited in \cite{Moore:1989vd}}
is \begin{equation}
{\rm ln}(Z) \simeq  (g-1) (N\sqd-1)~ {\rm ln}(k) + O(k\uu 0)\ .
\label{CSDegenHigherg}\end{equation}
To understand this formula, we can use semiclassical analysis
(see, \it e.g., \rm pg. 96 of \cite{Moore:1989vd})
 to determine
the leading large-$k$ behavior of the number of states.  For a compact phase space,
the number of quantum states is given, for small Planck
constant $\hbar$, to the volume of phase space in units of $\hbar$:
\begin{equation}
n\llo{\rm states} = {\rm (const.)} \cdot \frac
{{{\tt Vol}\llo{\rm phase~space}}}{{\hbar\uu{{\frac{\tt Dim.}{2}}} }} 
~ \big [ 1 + O(\hbar) \big ] \ ,
\end{equation}

For Chern-Simons theory in canonical quantization, the phase space is the moduli space 
${\cal M}\llo{G,g}$
of flat $G$-connections on the spatial slice $\Sigma\llo g$, and the Planck constant
$\hbar$ is proportional to $\frac{1}{k}$.   The volume of the moduli space of flat connections is $k$-independent, and its dimension \cite{Witten:1988hf} is 
\begin{equation}
{\tt Dim.}({\cal M}_{G,g}) = (2g-2) ~{\rm Dim.}(G)\ .
\end{equation}
Therefore the number of quantum states, in the large-$k$ limit, is
\begin{equation}
Z = n\llo{\rm states} = {\rm (const.)} \cdot k\uu {\hh {\tt Dim.}({\cal M}_{G,g}) } 
 \big [ 1 + O(k\uu{-1}) \big ]
\end{equation}
and the entropy is
\begin{equation}
{\rm ln}(Z) = (g-1)~(N\sqd - 1) ~ {\rm ln}(k) + O(k\uu 0)\ .
\label{GenusgCSEntropy}
\end{equation}
The coefficient of the ${\rm ln}(k)$ term does not depend on the numerical, $k$-independent
factor in the volume of ${\cal M}_{G,g}$, only on its volume.  This order $N\sqd$ entropy overwhelms the entropy of the matter.  This $N\sqd~ {\rm ln}(k)$
divergence of the entropy is striking, because it is larger than any gravitational contribution
to the entropy, which would scale at most as $\frac{1}{G_N} = N$.
 
\subsection{Degrees Of Freedom}
We want to emphasize that the divergent entropy at large $k$ is not attributable
to the nonpositive scalar curvature of the boundary in the case where the boundary
is $S\uu 1 \times \Sigma\llo g, ~g\geq 1$.  It is known that CFT partition functions on
such geometries need not be convergent, and the corresponding bulk instabilities have
been studied in some cases. \cite{Maldacena:1998uz, Seiberg:1999xz, Witten:1999xp}.  However
the large-$k$ divergence of the entropy in CSM theory cannot be
an artifact of vanishing or negative
scalar curvature, as the instability is not present in some cases where
the entropy is nonetheless still logarithmically divergent with $k$.
In the case of the critical model, for instance, the unstable direction of the scalars is 
always stabilized independently of $k$, by
the quartic interaction.

In the case of the free scalar or the critical scalar, the partition function on $S\uu 3$ is
stabilized by the conformal coupling but still displays a ${\rm ln}(k)$
divergence in the free energy \cite{Kac:1988tf,Klebanov:2011gs}, 
\begin{equation}
F = - {\rm ln} (Z\llo {S\uu 3}) \simeq + \frac{N(N-1)}{2} ~ {\rm ln}(k) + O(k\uu 0)\ .
\end{equation}
This comes entirely from the Chern-Simons sector, as the conformal coupling of the scalars
allows them to contribute only terms analytic in $k$.  The value of $F = - {\rm ln}(Z\llo{S\uu 3})$
for various conformal and superconformal field theories in three dimensions has been an
object of much recent study (\cite{Jafferis:2010un, Jafferis:2011zi, 
Closset:2012vg, Myers:2010tj}), particularly the investigation of the hypothesis that $F$ is a measure
of the number of degrees of freedom of the system that decreases along renormalization
group flows, analogously to the $c$ coefficient in two dimensions \cite{Zamolodchikov:1986gt}
or the $a$ coefficient in four dimensions \cite{Cardy:1988cwa, Komargodski:2011vj}.  (A 
general proof of the equivalence between entanglement entropy in a 3-dimensional CFT and its free energy on $S\uu 3$ has been presented in \cite{Casini:2011kv}.)  With
this interpretation, we see again that there
are of order $N\sqd~{\rm ln}(k)$ degrees of freedom in the Chern-Simons-matter system \footnote{The tension between the Vasiliev bulk interpretation and $N^{2}$ degrees of freedom has also been emphasized by Klebanov (private communication)} \cite{Klebanov:2011gs}, attributable to the topological sector.

\subsection{Light States In ABJM Theory}
There have been proposals \cite{Giombi:2011kc, YinSlides} to derive Vasiliev gravity as a limit
of the ABJ theory \cite{Aharony:2008gk}.
For Chern-Simons-matter theories with ultraviolet-complete string
 duals, this same large-k divergence
 on a torus is natural when interpreted in light of string- and M- theory.  We can
for instance compactify the ABJM model on $T\uu 2$ rather than $S\uu 2$ spatial
slices, and ask what the holographic duality predicts, qualitatively, for the entropy.

\llsk Without doing a fully controlled calculation, we simply observe that the total entropy of the 
AdS should be approximately extensive in the radial direction, and that the entropy
at every point in the radial direction is divergent in the limit $k\to\infty$ with $N$ large
but fixed.  At any point in the radial direction, there are new states due to the topology that become
light at large $k$, corresponding to membranes that wrap the Hopf fiber of the $S\uu 7
 / Z\llo k$, and one direction of the longitudinal $T\uu 2$.  At large $N$ these states
 are still very heavy, but at fixed $N$, however large, the proper energy of these states,
 at any fixed point in the radius, goes to
 zero at large $k$, because the size of the Hopf fiber is $1/k$ in 11-dimensional Planck units.
The fixed-$N$, infinite-$k$ entropy contributed by any point in the radial direction diverges,
and this is visible in every duality frame.  In the type IIA duality frame, the Hopf
fiber is invisible, having been turned into the M-direction, but the
AdS radius in string units is inversely proportional to $k$, at fixed $N$.  Therefore fundamental
strings wrapping a cycle of the longitudinal torus become light, and make a divergent contribution
to the entropy.
As the longitudinal torus shrinks further towards the infrared, we T-dual to type IIB and
the T-dual radius decompactifies.  In this duality frame, there is a divergent entropy due
simply to the decompactification of the emergent T-dual dimension.

\llsk We could also ask what is the entropy of $N$ M2-branes wrapped
on $T\uu 2$ and probing a $\IC\uu 4 / \IZ\llo k$
singularity in M-theory, without taking the near-horizon limit or taking the
back-reaction into account.  This is a different approximation, but also illuminating because
we see again a naturally emerging divergent entropy at large $k$.  Reducing on
the $T\uu 2$ from M-theory to type IIB, we transform the M2-branes into $N$ particles
each carrying one unit of Kaluza-Klein momentum on the T-dual direction.  Even restricting
ourselves to normalizable states that saturate the BPS bound in this framework, we see an
entropy that diverges at large $k$.  Each of
these particles can occupy any of $k$ massless twisted sectors of the orbifold, and still
saturate the BPS bound for a Kaluza-Klein momentum unit.  Since each of $N$ interchangeable
particles can inhabit one of $k$ possible states, the total degeneracy of such
quantum states gives a contribution to the partition function of
\begin{equation}
\Delta Z \gtrsim k\uu N / N!\ ,
\end{equation}
because the symmetry factor by which one divides is no more than $N!$.
This corresponds to a contribution to the entropy of
\begin{equation}
\Delta {\rm ln}(Z)  \gtrsim N ~ {\rm ln}(k) - {\rm ln}(N!) \simeq N ~ {\rm ln}(\lambda\uu{-1}) \ , 
\end{equation}
which is remarkably similar to the Chern-Simons degeneracy (\ref{CSDegenTorus}).  

This counting is most likely an underestimate.  Though interactions between particles
may in principle lift some of these BPS vacua, a massive perturbation lifting
the flat directions allows us to reduce to Chern-Simons theory in the unhiggsed
vacuum and compute the supersymmetric index.  This classical vacuum alone
contributes to the index with the full degeneracy of the pure Chern-Simons system on
the torus for $U(N) \times U(N)$ at level $k$.

\subsection{\texorpdfstring{$N^{2}$}{N2} Entropy}

The $N\sqd$ scaling of the partition functions on $S\uu 3$ and $S\uu 1 \times
\Sigma\llo g$ with $g\geq 2$ indicates difficulties for the interpretation of the CSM theory in terms
of Vasiliev gravity.  The four-dimensional Newton constant $G_N$ as inferred from
stress tensor correlators is of order $1/N\uu 1$ in units of the AdS scale,
rather than $1/N\sqd$, so the order $N\sqd$ entropy cannot be attributed to a gravitational
effect like a horizon entropy if $L_{AdS}/N$ is indeed the true Newton constant of the
theory.  
Entropies proportional to $N\sqd$ are characteristic of matrices.  Here we see that the
vectorlike holography of Chern-Simons-matter systems rediscovers its matrixlike character.
In terms of the proposal to complete Vasiliev gravity in terms of an open-closed topological
string theory \cite{Aharony:2011jz,TopologicalStringVasiliev, YinSlides}, the $N\sqd$ scaling of
the entropy is an
indication that the graviton should reside in the closed string, rather than open string sector,
of such a theory, in accordance with the principle that it is the gravitational force that must
always carry the largest entropy \cite{Bekenstein:1980jp} and weakest interaction
\cite{ArkaniHamed:2006dz} of any sector of a quantum gravitational theory.  Reconciling
this with the identification $G\llo N \propto 1/N$ apparently dictated by stress tensor
correlation functions is a challenge for any proposal such as \cite{TopologicalStringVasiliev}.

\subsection{Higher Genus And Hyperbolic Black Holes}
To understand the bulk geometry dual
to this spatial geometry, mod out the bulk, presented in hyperbolic slicing, by the action of a
 discrete group.  This is a valid operation in any gravity theory, including Vasiliev gravity.  The correspponding bulk geometry is  a ``zero mass" hyperbolic black hole\cite{Emparan:1999gf}.
 The boundary dual of the "zero mass" black hole corresponds to the Chern-Simons Matter system on the Riemann surface at the temperature $\frac{1}{2\pi R_{\rm curv}}$, where
 $R\llo{\rm curv}$ is the curvature radius of the spatial slices \cite{Emparan:1999gf, Horowitz:2009wm, Myers:2010xs}.
 The point of unbroken gauge symmetry in the matter theory is unstable due to the ${\tt Ricci}\llo 3 \phi^2$ coupling, but interaction terms stabilize the scalar vev.
 In the critical case, for example, the $\phi\uu 4$ coupling stabilizes the
 scalars, independent of $k$.   (For the "free" scalar theory, the theory is not in fact strictly free either, due to the $\phi\uu 6$ interaction \cite{Aharony:2011jz},
which stabilizes the zero mode.)  In this case, there are no singular shrinking cycles in the bulk gravitational metric to blame for the light states but there is a finite area black hole horizon.  As mentioned above the normal geometric horizon entropy $S \sim 1/G_N \sim N$ is insufficient to account for the $N^2$ entropy found in the boundary theory.

It seems likely that tensionless winding strings are again relevant in this case.  If
we fix a point in the AdS radial direction,
the density of winding string states grows exponentially
as a function of length \cite{McGreevy:2006hk, Milnor, Margulis},
 so that there 
is a Hagedorn density with transition temperature $T_H \propto \ell\llo 0 / \alpha^\prime$,
where $\ell\llo 0$ is the proper size of the longitudinal spatial slices.
In the zero-tension limit $\alpha^\prime\to\infty$, the Hagedorn temperature goes to zero.  
At arbitrarily low temperatures, the formal entropic contribution of the winding states
exceeds the contribution of their partonic constituents, 
signaling that  the string
thermodynamics should break down in favor of an order $N\sqd$ entropy counting the
constituents, perhaps crossing over to a horizon entropy involving the closed string $G_N \sim 1/N^2$.

\subsection{RG Flow}
Understanding the renormalization group flow
of the theory to pure Chern-Simons theory may be useful for understanding the holographic dynamics of CSM theory, including the order $N\sqd$ entropy and the ${\rm ln}(k)$ divergence.
For many 3-manifolds, the
holographic dual to pure Chern-Simons theory is understood in terms of the topological
string \cite{Gopakumar:1998ki}, including cases where an order $N\sqd$ free energy is present. 
 For the
case of $S^3$ for example, there is a well-controlled dual in terms of the topological
string on the resolved conifold, where the singular behavior of the $k\to \infty$ limit
arises from the vanishing of the complexified K\"ahler parameter of the blown-up $\IC\IP\llo 1$ base of the resolved conifold, leading to unsuppressed contributions of worldsheet instantons.

\subsection{\texorpdfstring{$T^{3}$}{T3} Modular Invariance Constraints}
 The motivation to consider coupling the matter CFT to
large-$k$ Chern-Simons theory was originally to take the limit $k\to \infty$, in order to
implement a projection to the singlet sector of the operator spectrum.
\cite{Giombi:2011kc, Shenker:2011zf}.
 Given the difficulties of promoting this construction to a fully well-defined local
quantum field theory, one might wonder whether there may be some construction
the singlet-projected matter QFT without any additional states, perhaps some kind of
BF theory.  We can answer this question in
the negative.  For the partition function on $T\uu 3$, there is a simple demonstration
\cite{ChannelCovariance} that
such a construction cannot exist at all, based on modular invariance.  Treating one of the
three cycles, say $\theta^3$ as the Euclidean time direction, the singlet-projected partition
function is computed by taking the full partition function with boundary conditions such that
the matter fields are periodic up to a particular group transformation $g \in G$
around the $\theta\uu 3$ cycle, and then averaging (\it not \rm summing!) over $G$.  This
procedure is the same regardless of the shape and size of the $T\uu 3$.  However
a consistent, local quantum theory must have the same partition function when quantized
in any "channel", \it i.e. \rm with respect to the Hamiltonian and Hilbert space defined
by any foliation of the manifold.  If we switch the roles of $\theta\uu 1$ and $\theta\uu 3$,
treating the former as Euclidean time and the latter as a spatial direction, then
the average over boundary conditions on $\theta\uu 3$ generates a partition function
that is not only asymmetric with the theory in the $\theta\uu 3$ channel, but does not
have any consistent Hilbert space interpretation in the $\theta\uu 1$ channel whatsoever.

\fakeindent That is, let 
\begin{equation}
Z[L_1,  L_3 ; g_3] = {\rm partition~function~with~radii~\it L_1~\rm and~\it L_3, \rm ~~and~periodicity~\it g_3\rm ~along~\theta\uu 3}\ .
\end{equation}
Suppose a local CFT exists such that the Hilbert space on any slice always contains just exactly
the singlet sector of the full matter theory, and nothing more.  
Then the partition function for the singlet sector on a spatial slice with radius $L_1$ 
at temperature $T = \frac{1}{L_3}$ is
\begin{equation}
Z_{\rm singlet}[L_1, L_3] = \frac{1}{|G|}~\sum_{g_3\in G}~Z[L_1,  L_3 ; g_3].
\label{SingletProjection}
 \end{equation}
(If we take a continuous group $G$ with a discrete one, the formula is the same
except that the sum is replaced with an integral and the cardinality $|G|$ of
$G$ is replaced with the Haar volume.)  But there are at least two things wrong with this possibility.  First, the formula is not invariant under $L\llo1 \leftrightarrow L\llo3$.  Secondly,
if we fix $L\llo 2, L\llo 3$ and take $L\llo 1\to \infty$, the partition function does \it not \rm 
take the form of a sum of exponentials of $L\llo1$ with positive integer coefficients,
which as a consistency condition of a quantum field theory of any kind, it must.

\llsk That is to say, if any kind of Hilbert space exists at all in the $\th\uu 1$ channel, 
then it must be possible to write the partition function in the form
\begin{equation}
Z_{\rm any~consistent~theory}[L_1, L_3] = \sum_{{\rm states~in~\th\uu 1-channel}}~\expp{- L_1~ E^{(\th\uu 1-{\rm channel})}}\ ,
\label{ConsistentForm}\end{equation}
  where the energies $E^{(\th\uu 1-{\rm channel})}$ may depend on $L_3$ but
  the coefficients are 1 (or another positive integer, if there are degeneracies).  However
  the partition function (\ref{SingletProjection}) is realized in the $\th\uu 1$
  channel as an average (as opposed to a sum) of partition functions with different
  periodicities along the spatial cycle $\th\uu 3$.  Therefore it \it cannot \rm have the form
  (\ref{ConsistentForm}), unless the ground state energy of the unprojected theory in
  the $\th\uu 1$ channel would be independent of the boundary condition $g\llo 3$, which is
  not the case in general, and certainly not for free bosons or
  fermions.    Therefore the coefficient of the leading exponential of
  $L\llo1$, which in a consistent quantum theory encodes the ground state degeneracy 
  in the $\th\llo1$ channel, is fractional for this theory, signaling the nonexistence of a
  Hilbert space of any kind in this channel, let alone a Hilbert space isomorphic to the
  one in the $\th\uu 3$ channel. 
  
   This argument most directly rules out the existence of a consistent partition function
   for $T\uu 2$ spatial slices, but the inconsistency cannot be confined to
   this case alone; the existence of cobordisms -- that is, smooth geometries interpolating
   between spatial slices of different topology -- define charge-conserving maps 
   Hilbert spaces on the torus and on higher-genus Riemann surfaces.  Thus, if a local
   CFT did exist that contained only the singlet sector of the original matter CFT
   on slices of higher genus, the path integral on the interpolating manifold could
   be used to define the singlet theory on $T\uu 2$ spatial slices as well; but we know that this
   theory can have no consistent definition.
   
    This situation is similar to the case of two-dimensional CFT with global
   symmetries, where the truncation of the theory to the singlet sector is
   not consistent with modular invariance unless twisted states are added to supplement
    the Hilbert space.  The number of states that must be added increases with the cardinality
    of the group, leading to an infinite entropy when the group is continuous.  The
    light states of the Chern-Simons sector at large $k$ can be identified as analogous
    to the plethora of low-lying twisted states that appear when one tries to construct a
    modular-invariant orbifold by a group with a cardinality $|G|$ that is going to infinity. 
    
 \subsection{Light States In the \texorpdfstring{$W_N$}{WN} Models and their Gravity Duals}
 The lower dimensional duality described by a $W_N$ boundary theory \cite{Gaberdiel:2010pz,Gaberdiel:2011wb,Gaberdiel:2011nt,Ahn:2011pv,Chang:2011mz,Papadodimas:2011pf} has light states with such an origin.  These states have been described in \cite{Gaberdiel:2011aa} as twisted states in a continuous $SU(N)$ orbifold in the boundary theory, involving flat connections like the ones relevant in Chern-Simons theory.  The authors of \cite{Castro:2011iw,Gaberdiel:2012ku} 
have interpreted thses states in the bulk by \ as due to ``conical excess" solutions.  Here these states appear directly on the $S^1$ spatial geometry and are necessary for a consistent modular invariant solution and hence for the finite temperature black hole dynamics.  The light states we consider are not necessary for the bulk thermal dynamics with $S^2$ boundary.  No hint of them, or of light strings that could wrap the $T^2$ are visible there \cite{Shenker:2011zf, Giombi:2011kc}.

\subsection{dS/CFT}
One area in which Vasiliev gravity has been applied has been to the study of 
holographic cosmology, through the dS/CFT correspondence.  A
nonunitary version of the Chern-Simons-matter theory, based on replacing the
scalar bosons with scalar fermions, has been proposed as a holographic dual
for Vasiliev gravity in de Sitter space in 4 dimensions  \cite{Anninos:2011ui}.
The topology-dependent divergence of
the partition function noted in this article may have relevance for the meaning of
this correspondence, particularly for any sort of probabilistic interpretation of it \cite{Maldacena:2002vr}.

\subsection{Supersymmetric Extensions}
It would be interesting to analyze supersymmetric extensions
of the matter-Chern-Simons
theory with various amounts of supersymmetry,
from the minimal (${\cal N} = 1$ or ${\cal N}=2$) case \cite{Schwarz:2004yj}
to the almost-maximal (${\cal N}= 5,6$) case of the general ABJ \cite{Aharony:2008gk}
and ABJM \cite{Aharony:2008ug} theories,
and the maximal (${\cal N} = 8$) case
of the $k=1$ ABJM theory.
The addition of supersymmetry introduces new technical issues (for instance, exactly
flat directions on the moduli space of vacua) while promising a greater
degree of control over quantum effects.  A Vasiliev-type gravity dual has also been proposed
for the supersymmetric Chern-Simons-matter theory \cite{YinSlides}.

\subsection{Decoupling}
There is a sense in which these light states decouple at $k=\infty$.   From (\ref{qmmass}) we see that at $k = \infty$ the holonomy becomes an infinitely low-frequency degree of freedom and hence does not move.  Scattering of KK scalar modes will not change its value.  This does not mean that these light states can be removed from the theory.  To remove them would
be to fix definite values for the holonomies on the cycles of the spatial
slice.  But fixing the holonomy on spatial slices of the CFT does not
define a sensible bulk theory of any kind: With such a definition, the $T^3$ partition function would not be modular invariant;  the limit as $k \rightarrow \infty$ of physical quantities like the hyperbolic black hole entropy would not be smooth; and the higher spin correlators would not be uniquely defined independent of the order $N$ parameters
by which the holonomies are characterized.

\subsection{Condensed Matter Applications}
Finally we should note that our results may be useful in analyzing condensed matter quantum hall systems where another set of degrees of freedom become light and changes the quantum hall dynamics.  For a recent example, see \cite{barkeshli:2012ja}.

\section{Acknowledgements}
We thank Xi Yin for suggesting this problem and for helpful discussions about it.   We also thank D. Anninos, T. Banks,  M. Barkeshli,  A.Basu, J. David, R. Gopakumar, D.Gross, D.Harlow, T. Hartman, S. Hartnoll, D. Jafferis,  I. Klebanov,  Z. Komargodski, S. Kachru, J. Maldacena, A. Maloney,  G. Moore,  A.Sen, E. Silverstein, D. Stanford, L. Susskind,  A. Tomasiello, G. Torroba and S. Wadia for valuable discussions and correspondence.  We would like to thank J. Maldacena for pointing out an error in the original version of this paper and for sharing a draft of \cite{Aharony:2012ns} with us.   We would also like to thank  Douglas Stanford for pointing out typos in the draft. The work of S.H. was supported by the World Premier International Research Center  Initiative, MEXT, Japan, and also by a Grant-in-Aid for Scientific Research (22740153)
from the Japan Society for Promotion of Science (JSPS). S.H. is also grateful to the Stanford SITP, and the theory group at Caltech for hospitality while this work was in progress.  The research of SB, JM, and SS is supported by NSF grant 0756174 and the Stanford Institute for Theoretical Physics. S.B is also grateful to Harish-Chandra Research Institute, India and ICTS, Bangalore for their hospitality when this work was in progress. S.B would also like to acknowledge all the participants of ICTS DISCUSSION MEETING ON STRING THEORY, especially Rajesh Gopakumar for valuable discussion. The research of JM is also partially supported by the Mellam Family Fellowship and the Stanford School of Humanities and Sciences Fellowship. J.M. and S.H. are also grateful to the Cosmology and Complexity conference at Hydra for their hospitality while this work was in progress.  SS would like to acknowledge the Aspen Center for Physics and 
NSF Grant 1066293 for hospitality and support during this project.

\appendix
\section{SU(2) gauge theory}
In the case of SU(2) the potential on the moduli space has the form,
\begin{equation}
V(A_{1}^{3},A_{2}^{3}) = -\frac{2}{\pi} \sum_{(m,n)\neq (0,0)} \frac{1}{(m^{2}+n^{2})^{\frac{3}{2}}} cos(\frac{mA_{1}^{3}+nA_{2}^{3}}{2})
\end{equation}
where the SU(2) flat connection has been parametrized as $A_{i} = \frac{1}{2}A_{i}^{3}\sigma^{3}$ where $\sigma^{3}$ is the diagonal Pauli spin matrix.
So the same argument as in the Abelian case shows that we can expand around the trivial flat connection. The action for the reduced quantum mechanical problem is given by, 
\begin{equation}
{S} = \frac{ k}{8{\pi}} \int dt \ Tr(A_{1}\frac{d}{dt} A_{2})  + \int dt \  [\frac{d\phi^{\dagger}}{dt}\frac{d\phi}{dt}  - {\phi}^{\dagger}A_{i}A_{i}{\phi}]
\end{equation}
where $\phi$ is a complex two component column vector transforming in the fundamental representation of the $SU(2)$. $A_{1}$ and $A_{2}$ are $2\times 2$ traceless Hermitian matrices transforming in the adjoint representation of $SU(2)$. 
With this the Hamiltonian becomes 
\begin{equation}
H = \pi_{1}^{*}\pi_{1} + \pi_{2}^{*}\pi_{2} + {\phi}^{\dagger}A_{i}A_{i}{\phi}
\end{equation}
where $\pi_{i} (\pi_{i}^{*})$ are canonical momenta conjugate to ${\phi}_{i}({\phi}_{i}^{*})$.
The Hamiltonian again factorizes in this case. We can write 
\begin{equation}
\phi^{\dagger}A_{i}A_{i}\phi = \frac{1}{4} \phi^{\dagger}\sigma^{a}\sigma^{b}\phi A_{i}^{a}A_{i}^{b} = \frac{1}{4} \phi^{\dagger}\delta^{ab}\phi A_{i}^{a}A_{i}^{b}= \frac{1}{2}\phi^{\dagger}\phi \ Tr(A_{i}A_{i})
\end{equation}
where $A_{i} = \sum_{a=1}^{3} \frac{1}{2}A_{i}^{a}\sigma^{a}$ and $\sigma^{a}$ are generators of SU(2). So the Hamiltonian becomes 
\begin{equation}
H = \pi_{1}^{*}\pi_{1} + \pi_{2}^{*}\pi_{2} + \frac{1}{2}\phi^{\dagger}\phi \ Tr(A_{i}A_{i})
\end{equation}
We can see that the Hamiltonian has a manifestly factorized form. The Lagrangian has a global SU(2) symmetry which is the remnant of the original gauge symmetry of the field theory. So in the quantum mechanics we should set this charge to zero. This is given by the constraint, 
\begin{equation}
\frac{ k}{8\pi}i[A_{1},A_{2}]^{a} = J_{0}^{a}
\end{equation}
where $J_{0}^{a}$ is the Noether charge which generates the SU(2) rotations of the scalars. There are operator ordering ambiguities associated with the definition of the charge $J_{0}^{a}$. We shall resolve these issues in the next section where we shall discuss the case of a general $U(N)$ gauge group.

 We shall first quantize the Hamiltonian and then implement the constraint by projecting onto SU(2) invariant states in the Hilbert space.

In the $SU(2)$ case the reduced quantum mechanics model has a larger symmetry which is $SU(2)\times SU(2)$. The fields transform as 
\begin{equation}
\phi \rightarrow U_{1}\phi,\ A_{i} \rightarrow U_{2}A_{i}U_{2}^{\dagger}
\end{equation} 
where $U_{1}$ and $U_{2}$ are constant $SU(2)$ matrices. The original global gauge invariance of the quantum mechanical model is the diagonal $SU(2)$.  Now one can compute the Noether charges corresponding to these symmetries and what one finds is that left-hand side and the right-hand sides of the constraint are the generators of the individual $SU(2)$ transformations. The constraint is the statement that the physical wave functions are invariant under the diagonal symmetry transformations. Since the Hamiltonian factorizes we can start with product wave functions where each factor transforms in some definite representation of the respective $SU(2)$'s. Then the Gauss's law constraint is satisfied by picking up the singlet in the product representation.

\section{Effective Potential}
The method we use in this appendix is similar to that used in \cite{Aharony:2005ew}, except that we use heat-kernel method and so can be easily generalized to the case when the spatial slice is any higher genus Riemann surface.

We want to compute the determinant of the operator $-D^{2}$ where 
\begin{equation}
D_{\mu} = \partial_{\mu} + iA_{\mu}
\end{equation}
and $A_{\mu}$ is a flat gauge field of $U(N)$ in the fundamental representation.  One way of doing this is to solve the heat equation for this operator. So let us start with a quantum mechanical problem in euclidean space. The Euclidean propagator is defined as ,
\begin{equation}
G(x',s; x,0) = <x'\exp(-sH)|x>
\end{equation} 
where $s$ is a fictitious Euclidean time and $H$ is the Hamiltonian. The propagator satisfies the boundary condition :
\begin{equation}
G(x',s; x,0) \rightarrow {\delta}(x'-x) 
\end{equation} 
as $s\rightarrow 0$. The equation satisfied by the propagator is the Euclidean wave equation (Heat equation)
\begin{equation}
-\frac{\partial}{\partial s} G(x',s;x,0) = H_{x'} G(x',s;x,0)
\end{equation}
Now we expand the propagator in terms of a complete set of eigenfunctions of the Hamiltonian:
\begin{equation}
G(x',s;x,0) = \sum_{n} exp(-sE_{n}) <x'|n><n|x>
\end{equation}
It follows from this expansion that
\begin{equation}\label{mainequation}
\int_0^\infty \frac{ds}{s}\int dx\, G(x,s;x,0) = \sum_{n}(-ln E_{n}) = - ln\prod_{n} E_{n} = - ln Det (H)
\end{equation}
This is our main equation\footnote{We have done the standard thing of subtracting the infinite constant $\int^{\infty}_{0}\frac{ds}{s}e^{-s}$ in the identity $\log{b} =\int^{\infty}_{0}\frac{ds}{s}(e^{-s} - e^{-sb})$ to obtain (\ref{mainequation}). The result is obtained by a standard  renormalization.}.
So we can compute the determinant of the operator $H$ if we know the propagator of the corresponding quantum mechanical problem.
\subsection{Gauge Theory}
In our case the Hamiltonian of the quantum mechanical problem is $H= -D^{2}$. So the Hamiltonian is an $N\times N$ matrix-valued differential operator.  As a result the propagator is also an $N\times N$ matrix. Under a gauge transformation the Hamiltonian transforms as:
\begin{equation}
H_{x}\rightarrow U(x) H_{x} U(x)^{-1} 
\end{equation}
where $U(x)$ is a $U(N)$ valued gauge transformation. In the case of matrix valued differential operators and matrix valued propagators the formula for the determinant has the following form,
\begin{equation}\label{determinant}
\int_0^\infty \frac{ds}{s}\int dx\, Tr G(x,s;x,0) = - ln Det (H)
\end{equation}
where the trace is over the internal matrix indices. The form of the heat equation remains unchanged and covariance under gauge transformations requires the propagator to transform as
\begin{equation}
G(x',s;x,0) \rightarrow U(x')G(x',s;x,0)U(x)^{-1}
\end{equation} 
It is important that the trace computed in (\ref{determinant}) at the coincident points, $x=x'$, is invariant under this gauge transformation. The gauge transformation function does not depend on the fictitious euclidean time $s$.
So we have to solve the quantum mechanical problem of a particle carrying isotopic spin moving in the background of flat non-abelian gauge field. Since the background gauge field is flat at least locally the answer has the form
\begin{equation}\label{flatanswer}
G(x',s;x,0) \approx  Pexp(-i \int_x^{x'} A) \\G_{0}(x',s;x,0)
\end{equation}
where $P$ is the path ordering symbol and $G_{0}$ is the free propagator in the absence of the gauge field. We have not specified any particular path for the Wilson loop because the connection is flat and we are looking at a local patch and so the path chosen for the Wilson loop is homotopically trivial. If the space\footnote{By space we mean the three dimensional space on which the Chern-Simons matter theory lives. The fictitious Euclidean time does not play any role in our discussion. } is simply connected then this answer is exact and it shows that the propagator evaluated at coincident points, $x=x'$, is the same as the free one. So the heat kernel formula tells us that the determinant evaluated with a background gauge field is the same as the free one. This also follows from the facts that a flat connection in a simply connected space can be gauged away by a non-singular gauge transformation and the determinant is gauge invariant under such a transformation. 

In (\ref{flatanswer}) $G_{0}$ is the free propagator and so it is proportional to the identity matrix in $U(N)$ space. Since it is the free propagator it does not participate in the gauge transformation and so the $G$ defined in (\ref{flatanswer}) has the correct gauge transformation property which follows from the gauge transformation property of the Wilson line. 

\subsection{Multiply Connected Space}
If the space is not simply connected then (\ref{flatanswer}) has to corrected. Since the configuration space of the particle is multiply connected we have to work on the simply connected covering space of the configuration space.  The propagator on the original configuration space can be derived from the covering space propagator by the method of images. This works because the heat equation is a linear first order PDE. So if we can write down a solution of the heat equation which satisfies the boundary condition then that is the unique solution.

Let us denote by $M$ the three dimensional space on which the gauge theory lives. $M$ is not simply connected. Let us denote the simply connected covering space by $\hat M$. So we can write:
\begin{equation}
M = \frac{\hat M}{\pi _{1}(M)}
\end{equation}where ${\pi _{1}(M)}$ fundamental group of $M$. This equation means that there is  discrete group $\mathcal{G}$ isomorphic to the fundamental group of $M$ which acts freely on $\hat M$ and $M$ is the quotient of $\hat M$ by the action of this group.  The universal cover $\hat M$ is unique modulo diffeomorphism. We shall also assume the following things. The universal cover $\hat M$ has a metric and the group $\mathcal{G}$ is a discrete subgroup of the group of isomorphisms of the metric. As a result of this the metric induced on the quotient $M$ is the same as the metric on the cover $\hat M$\footnote{We shall relax this condition by considering conformally coupled matter fields.}. Since the operator depends on the metric if we neglect the gauge field then the heat equations are the same on the base and the covering space. Now we have to lift the gauge fields to the covering space. Since gauge fields are well defined function on the base $M$ they lift to periodic functions on the cover, i.e, the lifted gauge fields satisfy the property that, $A(x) = A({\gamma}x)$, $\forall {\gamma}\in \mathcal{G}$ and $x\in \hat M$. So the gauge fields are constant on the orbits of $\mathcal{G}$ \footnote{Tis is not a gauge invariant statement on the covering space. We shall discuss this in the following section.}. The lifted gauge fields are just the pullback of the gauge fields on base to the cover by the covering map and they are also flat on the covering space. For example in the case of $T^{2}\times R^{1}$ the covering space is $R^{2}\times R^{1}$ and the group $\mathcal{G}$ is the group of discrete translations of the plane which are isometries. The torus is the quotient of the plane by the discrete translations. The flat gauge fields on the torus lift to flat gauge fields on the plane which are periodic on the lattice. the holonomy of the gauge field along $a$ and $b$ cycles become the values of the Wilson lines  of the gauge field along the two sides of the unit cell of the lattice.  

\subsection{Torus}
We first solve the problem for $T^{2}\times R^{1}$.  The coordinates on the torus are denoted by the complex numbers $(z,\bar z)$ with periodicity $z\sim z + 2{\pi}(m + n{\tau})$ where $(m,n)\in \mathbb Z$ and ${\tau}$ is the complex structure. The coordinate along $R^{1}$ will be denoted by $x$. The metric can be written as , ${ds}^{2} = dzd\bar z + {dx}^{2}$. Since the eigenfunctions of the Hamiltonian operator are periodic on the torus and the propagator can be written in terms of the eigenfunctions of the operator it satisfies the same periodicity property:
\begin{equation}\label{periodicprop}
 G(z',\bar z',x',s;z,\bar z,x,0) = G(z'+2{\pi}(m'+n'{\tau}),\bar z' +c.c,x',s;z+2{\pi}(m+n{\tau}),\bar z+c.c,x,0)
 \end{equation}
where $c.c$ stands for complex conjugate. So the propagator on ${T^{2}\times R^{1}}$ is periodic. Now we shall work on the covering space which in this case is the complex plane. The lattice is generated by the complex numbers $(2{\pi},2{\pi}{\tau})$ and can be identified with the Abelian group $\mathbb Z \oplus \mathbb Z \sim 2{\pi}\mathbb Z \oplus 2{\pi}{\tau}\mathbb Z$. $\mathbb Z\oplus \mathbb Z$ is precisely the homotopy group of the torus and we can write, $T^{2} = \frac{R^{2}}{\mathbb Z \oplus \mathbb Z} \sim \frac{\mathbb C}{2{\pi}\mathbb Z \oplus 2{\pi}{\tau}\mathbb Z}$. So the covering space of the total geometry $T^{2}\times R^{1}$ can be written as $R^{2}\times R^{1}$ and $T^{2}\times R^{1} = \frac{R^{2}}{\Gamma}\times R^{1}$, where ${\Gamma}$ is the lattice. 

The flat gauge field  has nontrivial holonomies associated with the two noncontractible cycles in the geometry associated with the torus factor. The gauge field can be lifted to the covering space and the lifted gauge field satisfies the periodicity condition, $A(z+(m,n),\bar z+(m,n),x) = A(z,\bar z,x)$, where $(m,n)$ is a lattice translation vector. The periodicity condition is not a gauge invariant statement on the covering space. But one can think of it as a partial fixing of gauge in the covering space. The gauge transformations which survive are precisely those that have the same periodicity as the lattice. But they are also the gauge transformations which descend to the base. So the group of allowed gauge transformations on the cover are the same as the group of allowed gauge transformations on the base after this partial "gauge fixing".  

We can also think of this in the following way. The unit cell of the lattice with its sides periodically identified is identical to the torus. So any geometrical object defined on the torus can be defined on a single unit cell without any change. Once the object is defined in a single unit cell it can extended to the whole lattice by imposing the periodicity condition. In the case of gauge field this can be thought of as a gauge fixing condition. Since the covering space is simply connected there are no nontrivial flat connections on the covering space and so we can gauge it away. But the gauge transformation that we have to make does not satisfy the periodicity condition and so is not an allowed gauge transformation. 

The propagator in the covering space is :
\begin{equation}\label{covingspaceprop}
\bar G(z',\bar z',x',s;z,\bar z, x, 0) = W(z',\bar z',x';z,\bar z,x) G_{0}(z',\bar z',x',s;z,\bar z, x, 0)
\end{equation}
where $G_{0}$ is the free propagator on the cover and is proportional to the identity matrix in the $U(N)$ space. $W$ is the same Wilson line that appears in (\ref{flatanswer}). Now let us consider the case where the points $(z,z')$ belong to the same unit cell of the lattice. In that case the propagator $\bar G$ is a potential candidate for the propagator on $T^{2}\times R^{1}$. But this propagator does not satisfy the periodicity condition eqn-(12). The correct propagator can be obtained by summing over the lattice translation vectors and can be written as :
\begin{align}\label{latprop}
 G(z',\bar z',x',s;z,\bar z, x, 0)&\ =\ \sum_{{\gamma}\in{\Gamma}}\bar G(z' +{\gamma},\bar z'+\bar{\gamma},x',s;z,\bar z, x, 0) \nonumber\\ 
 &= \sum_{{\gamma}\in{\Gamma}}W(z'+{\gamma},\bar z'+\bar{\gamma},x';z,\bar z,x) G_{0}(z'+{\gamma},\bar z'+\bar{\gamma},x',s;z,\bar z, x, 0)
 \end{align}
 where $(z,z')$ belong to the same unit cell of the lattice. This is the propagator on $T^{2}\times R^{1}$. This satisfies the condition (\ref{periodicprop}). In the above formula the summation is only over the final position of the propagator. The same answer can be obtained by summing only over the initial position of the propagator. The reason for this is the following. The propagator $\bar G$ satisfies the condition that 
\begin{equation}
\bar G(z' +{\gamma},\bar z'+\bar{\gamma},x',s;z+{\gamma},\bar z + \bar{\gamma}, x, 0) = \bar G(z',\bar z',x',s;z,\bar z, x, 0),  \forall{\gamma}\in{\Gamma}
\end{equation}
This follows from the periodicity of everything under consideration along with the fact that the discrete group acts on the covering space as a group of isometries. As a result every unit cell is isometric to every other. So this result is almost trivial. So we have the following identity,
\begin{equation}
\bar G(z' +{\gamma}',\bar z'+\bar{\gamma}',x',s;z+{\gamma},\bar z + \bar{\gamma}, x, 0) =  \bar G(z' +{\gamma}' -{\gamma},\bar z'+\bar{\gamma}' - \bar{\gamma},x',s;z,\bar z, x, 0)
\end{equation}
Using this we can reduce any double sum over both the final and initial points can be reduced to a single sum over either the initial point or the final point.

Now we shall show that this also satisfies the correct boundary condition as $s\rightarrow 0$. As $s\rightarrow 0$ every term in the summation of (\ref{latprop}) is proportional to a delta function, ${\delta}^{2}(z' + {\gamma}-z){\delta}(x'-x)$, which comes from the free propagator $G_{0}$. Now according to our assumption the pair $(z,z')$ in (\ref{latprop}) refer to two points in the same unit cell. So the pair of points $z$ and $z'+{\gamma}$ can never coincide unless ${\gamma}=0$. So in the limit $s\rightarrow 0$, the only term which survives is ,$W(z',\bar z',x';z,\bar z,x) G_{0}(z',\bar z',x',s;z,\bar z, x, 0) = W(z',\bar z',x';z,\bar z,x){\delta}^{2}(z' -z){\delta}(x'-x)={\delta}^{2}(z' -z){\delta}(x'-x)$. So our solution satisfies the correct boundary condition.  

\subsection{Trace of The Propagator at Coincident Points }
To compute the determinant of the operator we have to compute the trace of the propagator at coincident points. This is the quantity $trG(z,\bar z,x,s;z,\bar z,x,0)$, where $tr$ is over the internal gauge indices. In our case the free propagator is proportional to the identity matrix where the proportionality constant is just the heat kernel of a single free complex scalar field in the covering space $R^{2}\times R^{1}$. We denote this quantity by the same symbol $G_{0}$. So we can write,
\begin{equation}
TrG(z,\bar z,x,s;z,\bar z,x,0) = \sum_{{\gamma}\in{\Gamma}}G_{0}(z+{\gamma},\bar z+\bar{\gamma},x,s;z,\bar z, x, 0)  TrW(z+{\gamma},\bar z+\bar{\gamma},x;z,\bar z,x) 
\end{equation}
The trace over the Wilson line can be expressed in terms of the trace of the products of holonomies along the $a$ and $b$ cycle in the following way. First of all the trace of the Wilson loop is independent of the choice of the point $(z,\bar z,x)$. This can be proved in the following way. 

We shall prove this in the general case where the group $\mathcal{G}$ acting on the covering space is nonabelian. This will be the case if say the spacetime manifold has the geometry ${\Sigma}_{g}\times R^{1}$ where ${\Sigma}_{g}$ is a genus $g$ Riemann surface with $g\geq 2$. In this case the the flat gauge fields are genuinely non-abelian. So let us consider the Wilson line $W({\gamma}x,x)$ where ${\gamma}\in \mathcal{G}$. $x$ is an arbitrary point on the covering space and ${\gamma}x$ is its image under the action of ${\gamma}$. In the case of torus ${\gamma}x$ is the translation of $x$ by a lattice translation vectors.

The Wilson line $W({\gamma}x,x)$ goes from $x$ to ${\gamma}x$. Let us choose another pair of points $(x',{\gamma}x')$ and consider the Wilson line $W({\gamma}x',x')$. Now let us  consider the four paths $\overrightarrow{(x,{\gamma x})}$, $\overrightarrow{({\gamma x},{\gamma}x')}$, $\overrightarrow{({\gamma}x',x')}$ and $\overrightarrow{(x',x)}$. They form a closed path and the shape of each path is arbitrary as the connection is flat. The holonomy along this closed path on the cover is zero and so we can write,
\begin{equation}
W(x,x')W(x',{\gamma}x')W({\gamma}x',{\gamma}x)W({\gamma}x,x) =1
\end{equation}
Now due to the periodicity condition,  $A(x)=A({\gamma}x)$, satisfied by the gauge field on the cover we have $W({\gamma}x',{\gamma}x)=W(x',x) = W^{-1}(x,x')$. So the zero holonomy condition reduces to 
\begin{equation}
W(x,x')W(x',{\gamma}x')W^{-1}(x,x')W({\gamma}x,x) =1
\end{equation}
This gives us 
\begin{equation}
W({\gamma}x,x)=W(x,x')W({\gamma}x',x')W^{-1}(x,x')
\end{equation}
So $TrW({\gamma}x,x)=TrW({\gamma}x',x')$. This proves our claim.

We can see form the above equation that in the Abelian case the Wilson line itself is an invariant quantity. But in the non-abelian case the Wilson lines are related by a conjugation and so the $Tr$ is an invariant quantity.

Now we again consider the case of the torus. Since the trace is independent of $(z,\bar z,x)$ we can choose a convenient value for the coordinates. Let us choose $(z,\bar z,x)$ to be the center $(0,0,0)$ and draw the lattice such that the center coincides with one vertex of a unit cell. So the the Wilson line $W(z+{\gamma},\bar z+\bar{\gamma},x;z,\bar z,x)$ becomes $W({\gamma},\bar{\gamma},0;0,0,0) $. Since the coordinate along $R^{1}$ does not play any role and the Wilson line is path-independent we can choose a path which lies on the slice $x=0$. So for ${\gamma}= m\vec a+ n\vec b$ where $(m,n)\in\mathbb Z\oplus \mathbb Z$ and $(\vec a,\vec b)$ are a set of basis vectors for the lattice, we can write$ W({\gamma},\bar{\gamma},0;0,0,0) = W(m\vec a+n\vec b)= W(a)^{m}W(b)^{n}$. $W(a)$ and $W(b)$ are the values of the Wilson line along the two sides of a unit cell. Since the sides of a unit cell represented by $(\vec a,\vec b)$ get mapped to the $a$ and $b$ cycles of the torus, $W(a)$ and $W(b)$ are precisely the holonomy of the flat connection along the two cycles of the torus. In this case they are all Abelian and so there is no ordering ambiguity. So the determinant can be written as,
\begin{equation}
-\ln Det(-D^{2}) = \sum_{(m,n)\in \mathbb Z\oplus\mathbb Z} A(m\vec a+n\vec b)Tr[W(a)^{m}W(b)^{n}]
\end{equation}
where $A(m\vec a +n\vec b)$ is given by the equation
\begin{equation}
A(m\vec a+n\vec b) = \int_{\epsilon}^\infty \frac{ds}{s}\int_{T^{2}\times R^{1}} dxdzd\bar z\,G_{0}((z,\bar z)+m\vec a+ n\vec b,x,s;(z,\bar z),x,0)
\end{equation} 
We have introduced a UV-cutoff ${\epsilon}$ in the $s$ integral. The expression for $G_{0}$ is known. The effective action obtained by integrating out the scalar field is given by $\ln Det(-D^{2})$. Now the heat-kernel of the free laplacian on $R^{d}$ is given by :
\begin{equation}\label{heatKfreeL}
G_{0}(x,s;y,0)= \frac{1}{(4{\pi}s)^{\frac{d}{2}}} exp(-\frac{(x-y)^{2}}{4s})
\end{equation}
Using this we get :
\begin{equation}
A(m\vec a+n\vec b) = \frac{vol(T^{2}\times R^{1})}{\pi} \frac{1}{|m\vec a+n \vec b|^{3}}
\end{equation}
So the final answer for the effective action is :
\begin{equation}
-S_{eff} = \frac{vol(T^{2}\times R^{1})}{\pi}\sum_{(m,n)\neq (0,0)}\frac{1}{|m\vec a+n \vec b|^{3}}Tr[W(a)^{m}W(b)^{n}]
\end{equation}
When the scalar field has a mass, $M$, the relevant operator is, $-D^{2} + M^{2}$ and we have to calculate $Det(-D^{2} + M^{2})$. The eigenvalues of the new operator is related to new by, $\lambda_{new} = \lambda_{old} + M^{2}$. So form equation $(A5)$ we conclude that,
\begin{equation}
G_{(-D^{2} + M^{2})} = e^{-M^{2}s}G_{(-D^{2})}
\end{equation}
Using this relation one can show that the effective action in the massive case is given by,
\begin{equation}
-S_{eff} = \frac{1}{\sqrt 2} \frac{vol(T^{2}\times R^{1})}{\pi^{\frac{3}{2}}}\sum_{(p,q)\neq(0,0)} \frac{M^{\frac{3}{2}}K_{\frac{3}{2}}(M|p\vec a + q\vec b|)}{|p\vec a + q\vec b|^{\frac{3}{2}}} Tr(W(a)^{p}W(b)^{q})
\end{equation}
where $K_{\frac{3}{2}}(\alpha)$ is the modified Bessel function of the second kind of order $\frac{3}{2}$. The leading asymptotic behavior of the function for $\alpha>> 1$ is given by,
\begin{equation}
K_{\frac{3}{2}}(\alpha) \sim \sqrt{\frac{\pi}{2\alpha}} e^{-\alpha}
\end{equation}  
So in the limit where $MR>>1$, the effective action can be approximated by,
\begin{equation}
-S_{eff} = \frac{vol(T^{2}\times R^{1})}{2\pi}\sum_{(p,q)\neq(0,0)} \frac{M e^{-M|p\vec a + q\vec b|}}{|p\vec a + q\vec b|^{2}} Tr(W(a)^{p}W(b)^{q})\end{equation}
$R$ is the size of the torus.
\subsection{Modular Invariance}
The effective action has been expressed in terms of the holonomy of flat connection along the two cycles of the torus, which corresponds to a particular choice of a set of basis vectors for the lattice. But the choice of cycles or the two basis vectors of the lattice is not unique. Any two choices are related by a $SL(2,\mathbb Z)$ transformation. So any automorphism of the lattice is a symmetry of the effective action. 

The $R^{1}$ part of the geometry does not play any role in the discussion. So we shall denote by $(\vec a,\vec b)$ the basis vectors of the lattice on some particular slice at some arbitrary value of $x$. We can choose a different set of basis vectors denoted by $(\vec a',\vec b')$ which are related to the old basis by,
\begin{align}\label{trans}
\vec a' = p\vec a + q\vec b \nonumber\\
 \vec b' = r\vec a + s\vec b
\end{align}
where $(p,q,r,s)$ are integers and satisfy, $ps-qr = 1$. So this is a $SL(2,\mathbb Z)$ transformation. 
Now the effective action can be written as 
\begin{equation}\label{seff}
-S_{eff} = \sum_{(m,n)\in \mathbb Z\oplus\mathbb Z} A(m\vec a + n\vec b)Tr[W(a)^{m}W(b)^{n}] 
= \sum_{(m,n)\in \mathbb Z\oplus\mathbb Z} A(m\vec a' + n\vec b')Tr[W(a')^{m}W(b')^{n}] 
\end{equation}
where $(\vec a',\vec b')$ are related to $(\vec a,\vec b)$ by  transformation (\ref{trans}). The new holonomies are related to the old ones by:
\begin{align}
W(a') = W(a)^{p}W(b)^{q}\nonumber\\
W(b')= W(a)^{r}W(b)^{s}
\end{align}
It is easy to check the equality in ({\ref{seff}) using these identities. So the effective action is modular invariant. This $SL(2,\mathbb Z)$ should give rise to Ward identities.

\subsection{ \texorpdfstring{$T^{2}\times S^{1}$}{T2S1}}
In this section we shall write down the answer when the field theory lives on $T^{2}\times S^{1}$. In this the cover is again $R^{2}\times R^{1}$. The group $\mathcal{G}$ is now $\mathbb Z\oplus \mathbb Z\oplus \mathbb Z$. The lattice is three dimensional with basis vectors denoted by $(\vec a,\vec b,\vec c)$. The metric is the flat metric. Since the homotopy group is $\mathbb Z\oplus \mathbb Z\oplus \mathbb Z$, the holonomies are commuting. The determinant is given by:
\begin{equation}
-lnDet(-D^{2}) = \sum_{(m,n,p)\in \mathbb Z\oplus\mathbb Z\oplus \mathbb Z} A(m\vec a+n\vec b+p\vec c)Tr[W(a)^{m}W(b)^{n}W(c)^{p}]
\end{equation}
where $A(m\vec a+n\vec b+p\vec c)$ is given by:
\begin{equation}
A(m\vec a+n\vec b+p\vec c) = \int_{\epsilon}^\infty \frac{ds}{s}\int_{T^{2}\times S^{1}} d^{3}x \,G_{0}(\vec x +m\vec a+n\vec b+p\vec c,s;\vec x,0)
\end{equation} 
where $G_{0}$ is the heat propagator that appears in the case of $T^{2}\times R^{1}$. In this case the group of automorphisms of the lattice is
 $SL(3,\mathbb Z)$ and so the effective action has this symmetry. This should also give rise to Ward identities.
 
 The final answer for the effective action in this case can be written as :
\begin{equation}
-S_{eff} = \frac{vol(T^{2}\times S^{1})}{\pi}\sum_{(m,n,p)\in\mathbb Z\oplus\mathbb Z\oplus \mathbb Z}\frac{1}{|m\vec a+n \vec b+p \vec c|^{3}}Tr[W(a)^{m}W(b)^{n}W(c)^{p}]
\end{equation}
\subsection{General Case}
In the general case the answer is:
\begin{equation}\label{logdet}
-\ln Det(-D^{2}) = \sum_{{\gamma}\in \mathcal{G}} A({\gamma})TrW({\gamma}\vec 0,\vec 0)
\end{equation}
where $A({\gamma})$ is given by,
\begin{equation}
A({\gamma}) = \int_{\epsilon}^\infty \frac{ds}{s}\int_{{\Gamma}} G_{0}({\gamma}x,s;x,0)
\end{equation}
In the general case the holonomies are non-Abelian. We have already proved that the $TrW({\gamma}x,x)$ is independent of the choice of $x$ even in the non-Abelian case. So in writing down the formula we have chosen an arbitrary value for $x$ which we have denoted by $\vec 0$. ${\gamma}\vec 0$ is the image of this point under the action of ${\gamma}\in \mathcal{G}$. In general the lattice has to be drawn on a space other than $R^{n}$. For example if we are working on the space-time geometry ${\Sigma}_{g}\times R^{1}$ where ${\Sigma}_{g}$ is a genus $g$ Riemann surface, then the lattice has to be drawn on the disc$\times R^{1}$ with hyperbolic metric on the disk. The metric on the space-time can be taken to be:
\begin{equation}
ds^{2} = \frac{4|dz|^{2}}{(1-|z|^{2})^{2}} + dx^{2}
\end{equation}
where $z$ is the coordinate on the unit disc. The free propagator has to be evaluated on this space. This answer can be found in literature. The domain of integration ${\Gamma}$ is the fundamental region for the action of group $\mathcal{G}$ on the covering space. This can be identified with a unit cell of the lattice    times whatever simply-connected non-compact direction the geometry has. 

We can choose the the point $\vec 0$ to be one of the vertices of the lattice in the same way as we did in the case of torus. The trace of the Wilson line on the covering space can again be expressed in the same way except that now the ordering has to be maintained. 

\subsection{A Short Proof Of (\ref{covingspaceprop})}
Heat equation on the covering space has the form:
\begin{equation}
-\frac{\partial}{\partial s} G(x',s;x,0) = -D_{x'}^{2} G(x',s;x,0)
\end{equation}
where $D_{{\mu}} = \partial_{\mu} + iA_{{\mu}}$ . Now one can write:
\begin{equation}
D_{{\mu}} = W(x)\partial_{{\mu}} W^{-1}(x)
\end{equation} 
where $W(x)=Pexp(-i\int_{x_{0}}^{x} A)$. $x_{0}$ is an arbitrary initial point and we have not specified any path for the integration because the connection is flat. On the simply connected covering space $W(x)$ is a well-defined function of $x$. Using (\ref{logdet}) it is easy to see that $W^{-1}(x')G(x',s;x,0)$ is the heat kernel of the free Laplacian $-{\partial}^{2}$. Therefore
\begin{equation}
G(x',s;x,0) = W(x')G_{0}(x',s;x,0)= Pexp(-i\int_{x_{0}}^{x'} A)K_{0}(x',s;x,0)
\end{equation}
Now the boundary condition as $s\rightarrow 0$ requires us to choose $x_{0} =x$. This concludes our proof.

\newpage
\bibliographystyle{JHEP}
\bibliography{Bibliography}
\end{document}